\documentclass{article}
\usepackage[affil-it]{authblk}

\usepackage{float}
\usepackage[a4paper, total={6in, 8in}]{geometry}

\usepackage{siunitx} 
\usepackage{graphicx} 
\usepackage{wrapfig}
\usepackage{amsmath} 
\usepackage{mathtools}
\usepackage{subfigure}
\usepackage{color}
\usepackage{enumitem}
\usepackage[colorlinks]{hyperref} 
\usepackage{chngcntr}
\counterwithin{figure}{section}
\usepackage[backend=bibtex,style=authoryear,natbib=true]{biblatex} 
\title{Estimating Shortest Path Length Distributions via Random Walk Sampling} 

\author{Minhui \textsc{Zheng} and Bruce D. \textsc{Spencer}} 

\affil{Department of Statistics, Northwestern University}

\date{\today} 

\begin{document}

\maketitle 

\begin{abstract}
In a network, the shortest paths between nodes are of great importance as they allow the fastest and strongest interaction between nodes. However measuring the shortest paths between all nodes in a large network is computationally expensive. In this paper we propose a method to estimate the shortest path length (SPL) distribution  of a network by random walk sampling. To deal with the unequal inclusion probabilities of dyads (pairs of nodes) in the sample, we generalize the usage of Hansen-Hurwitz estimator and Horvitz-Thompson estimator (and their ratio forms) and apply them to the sampled dyads. Based on theory of Markov chains we prove that the selection probability of a dyad is proportional to the product of the degrees of the two nodes. To approximate the actual SPL for a dyad, we use the observed SPL in the induced subgraph for networks with large degree variability, i.e., the standard deviation is at least two times of the mean, and for networks with small degree variability, estimate the SPL using landmarks for networks with small degree variability. By simulation studies and applications to real networks, we find that 1) for large networks, high estimation accuracy can be achieved by using a single random or multiple random walks with total number of steps equal to at least $20\%$ of the nodes in the network; 2) the estimation performance increases as the network size increases but tends to stabilize when the network is large enough; 3) a single random walk performs as well as multiple random walks; 4) the Horvitz-Thompson ratio estimator performs best among the four estimators.
\end{abstract}

\section{Introduction}
In a large network, the shortest paths between nodes are of particular importance because they are likely to provide the fastest and strongest interaction between nodes (\cite{katzav2015analytical}). Although measures such as diameter and mean distance (\cite{newman2010networks}, \cite{chung2002average}, \cite{cohen2003scale}) have been studied extensively, the entire shortest path length distribution (SPLD) has received little attention. While the shortest path for a pair of nodes is measurable by existing algorithms such as breadth-first search, measuring the shortest paths for all pairs of nodes in a large network is computationally expensive (\cite{potamias2009fast}).\\

In this paper, we study the problem of estimating SPLDs in networks via random walk sampling. In particular, for each possible value of the shortest path length (SPL), we estimate the fraction of dyads with that value of SPL. There are two aspects to the problem. First, if a dyad is observed in the sample, the observed SPL in the sample may exceed the actual SPL in the population. Second, the dyads observed in a random walk sample have unequal chances of being included in the sample. With regard to the former aspect, \cite{ribeiro2012multiple} have shown that in a network with large degree variability, random walks often uncover the shortest paths. In other words, for two nodes in a network where the variance of degree distribution is very large, the observed shortest path in the subgraph induced by a random walk sample is usually the true shortest path in the population. This property is present in scale-free networks where the degree distribution follows the power law. In this paper, we've shown that this property extends to networks whose degree distribution has a large coefficient of variation ($c.v.$), i.e., whose ratio of standard deviation to mean is large. On the other hand, \cite{potamias2009fast} have shown that in large networks, when calculating the actual distance is computationally expensive, one can use precomputed information to obtain fast estimates of the actual distance in very short time. More specifically, one can first choose a small fraction of nodes as landmarks and compute distances from every node to them. When the distance between a pair of nodes is needed, it can be estimated quickly by combining their precomputed distances to the landmarks.\\

With regard to dyads' unequal probabilities of being included in the sample, we draw upon classical sampling theory for estimating totals from samples of elements included with unequal probabilities. The estimators we use are Hansen-Hurwitz estimator (\cite{hansen1943theory}) and Horvitz-Thompson estimator (\cite{horvitz1952generalization}). Both estimators will be used in original form and ratio form to estimate the fraction of dyads with a particular value of SPL. The ratio form is defined with the numerator equal to the estimator of the number of dyads with a particular value of SPL and the denominator equal to the estimator of the total number of dyads. To develop the Hansen-Hurwitz estimator, we derive from theory of Markov chains (\cite{newman2010networks}, \cite{sigman2009}, \cite{anderson1989second}) that the expected number of appearances of a dyad in a random walk sample with a sufficiently large number of steps is approximately proportional to the product of the degrees of the two nodes. This result allows application of the Hansen-Hurwitz estimator to the sample including a duplicate selection of nodes. To develop the Horvitz-Thompson estimator, we approximate the random walk sampling of nodes by an adjusted multinomial sampling model in $t$ draws, with $t$ equal to the number of steps in the random walk. Then we apply the Horvitz-Thompson estimator to the sample excluding duplicate nodes.\\

We provide practical solutions to estimate $c.v.$'s and weights used in both estimators when we are only able to crawl part of the network and observe the actual degrees of the sampled nodes. We also provide plots and numerical measures to evaluate the performance of our estimators. By applying the estimators and evaluation techniques to several simulation studies, we have the following findings:

\begin{itemize}
\item When a network has a $c.v.$ in degree distribution much larger than $2$, random walks have strong ability to discover the actual shortest paths between sampled nodes. Therefore we can use the observed SPL between sampled nodes in the induced subgraph to approximate their actual SPL. 

\item When a network has a $c.v.$ in degree distribution much smaller than $2$, random walks don't have strong ability to discover the actual shortest paths between sampled nodes. Therefore we need to do breadth-first search in the population graph to get the actual SPL, but only to a fraction, such as $30\%$, of the sampled nodes (known as "landmarks"), and use that information to approximate the SPL between other sampled nodes. 

\item The estimation performance improves as sampling budget increases, with dramatic improvement as the sampling budget reaches $20\%$ and moderate improvement beyond that.

\item If we fix the total sampling budget, such as $20\%$, using a single random walk performs equally well as using multiple random walks.

\item To a small degree, the Horvitz-Thompson ratio estimator outperforms the generalized Hansen-Hurwitz ratio estimator, and people can use the former with a smaller sampling budget to achieve the same estimation accuracy as by the latter.

\item The estimation performance improves as the network size increases, but tends to be stable once the network is large enough, such as of size $n=5000$ or larger.
\end{itemize}

Finally, we apply our estimators to eight real networks with various sizes, degree distributions, and $c.v.$'s. The results from evaluation measures for estimation from real networks support our findings from the simulation studies. 

\section{Background}
\subsection{Preliminary Definitions}
Let $G=(V, E)$ be a finite graph (network), where $V$ is the set of nodes with $|V|=n$ and $S$ is the set of edges with $|E|=m$. Let $i\in \{1,...,n\}$ denote a node in the graph, and $r \in \{1, ..., N\}$ denote dyad $(i,j)$, $i, j=1,...,n$, $j\neq i$, in the graph, where $N=\binom{n}{2}$ is the number of dyads in the graph. An \textit{induced subgraph} $G^*=(V^*, E^*)$ of $G$, is a graph formed from a subset of the nodes $V^* \subset V$ and all of the edges $E^* \subset E$ connecting pairs of nodes in $V^*$.\\

The \textit{adjacency matrix $\boldsymbol{A}$} (\cite{newman2010networks}, p.111) of a graph is the matrix with element $A_{ij}$
such that 
\[
  A_{ij}=\left\{
                \begin{array}{ll}
                  1 \text{ if there is an edge from node $i$ to node $j$,}\\
                  0 \text{ otherwise.}
                \end{array}
              \right.
\]
A graph is \textit{undirected} if $A_{ij}=A_{ji}$ for all $i$ and $j$, i.e., the adjacency matrix $\boldsymbol{A}$ is symmetric. In this paper, we only consider undirected networks without self-edges, so the adjacency matrix is symmetric and the diagonal elements are all zero.\\

The \textit{degree} (\cite{newman2010networks}, p.133) of node $i$, denoted as $k_i$, in a graph is the number of edges connected to it. For an undirected graph, the degree can be written in terms of the adjacency matrix as
\begin{equation}
k_i = \sum_{j=1}^{n}A_{ij} = \sum_{j=1}^{n}A_{ji}
\end{equation}
We define $p_k$ to be the fraction of nodes in the network to have degree $k$, and the \textit{degree distribution} to be the collection of the $p_k$'s for $k=0, 1, ..., n-1$. We denote $<k>$ as the first moment and $<k^2>$ as the second moment of the degree distribution.\\

A \textit{path} (\cite{newman2010networks}, p.136) in a network is any sequence of nodes such that every consecutive pair of nodes in the sequence is connected by an edge in the network. A graph is \textit{connected} if and only if there exists a path between any pair of nodes.  A graph is \textit{primitive} if $A^k>0$ for some positive integer $k<(n-1)n^n$. In a primitive graph, a path of length $k$ exits between every pair of nodes for some positive integer $k$. The \textit{length} (\cite{newman2010networks}, p.136) of a path in a network is the number of edges traversed along the path. The \textit{shortest path} (\cite{newman2010networks}, p.139), also known as \textit{geodesic path}, is a path between two nodes such that no shorter path exists. The \textit{diameter $L$} of a graph is the longest shortest path between any two nodes. Note that the diameter is finite for connected graphs. \\

Let $l_r=l_{ij} \in \{1, ..., L\}$ denote the true \textit{shortest path length (SPL)}, also known as the \textit{geodesic distance}, of dyad $r$ in the population graph $G$. The \textit{mean distance $M$} of a graph, is the average of shortest path lengths of all dyads in the graph. We define $f_l$ to be the fraction of dyads in the network to have SPL $l$, and the \textit{Shortest Path Length Distribution (SPLD)} to be the collection of $f_l$'s for $l=1, 2, ..., L$

\subsection{Random Walk Sampling}
Random walk sampling is a class of network sampling methods that have arisen recently and has been applied widely in large networks, due to its strong ability of `crawling' in the network. In this paper, we define a single random walk $\{X_t\}$ with length $t$ ($t$ steps) in a given graph $G=(V,E)$ as follows:

1) Select a node $u$ with equal probability $1/n$ from $V$;

2) If node $u$ has $k_u$ neighbors, i.e., node $u$ has degree $k_u$, include one of its neighbors, say $v$, with equal probability $1/k_u$ into the sample;

3) In turn, conditionally independent of previous steps, one of $v$'s neighbor nodes is selected with equal probability $1/k_v$ from the set of $v$'s neighbors;

4) Repeat this process until the desired length $t$ of the random walk is reached.\\

In the real world, some random walks are self-avoiding, in which case an edge or a node cannot be visited twice. However, in this paper we only consider random walks that are allowed to go along edges more than once, visit nodes more than once, or retrace their steps along an edge just traversed. In other words, we may have duplicates in our random walk sample.

\subsection{Scale-free Networks }

Many of the research papers in graph theory concern the Erd\H{o}s-R\'enyi random graphs. A \textit{Erd\H{o}s-R\'enyi random graph} $G(n, p)$ is a graph with $n$ nodes and each edge is assigned independently to to each pair of distinct nodes with probability $p\in (0,1)$ (\cite{Kolaczyk2009:SAN:1593430}, p.156). By this definition, the degree distribution for a Erd\H{o}s-R\'enyi random graph follows a binomial distribution:
\begin{equation}
p_k=\binom{n-1}{k}p^k(1-p)^{n-1-k}.
\end{equation}
As demonstrated by (\cite{newman2010networks}, p.402), in the limit of large $n$, $G(n,p)$ has a Poisson degree distribution:
\begin{equation}
\lim_{n\rightarrow\infty}p_k=e^{-c}\frac{c^k}{k!},
\end{equation} 
where $c=(n-1)p$ is the mean degree of $G(n,p)$. According to the property of Poisson distribution, the variance of degree distribution is always equal to the mean of degree distribution.\\

The model is widely studied because of its simple structure. However, recent empirical results (\cite{albert2002statistical}) show that for many real-world networks the degree distribution significantly deviates from a Poisson distribution. In particular, for many real-world networks, the degree distribution has a power-law tail
\begin{equation}
p_k \propto k^{-\alpha},
\end{equation}
where $\alpha$ is the \textit{exponent} of the power law. Such networks are called \textit{scale-free}. Typically, the values in $\alpha$ from real networks are in range $[2,3]$, although values slightly outside this range are possible and are observed occasionally (\cite{newman2010networks}, p.248).\\

Scale-free networks possess some unusual properties as compared to other networks. One of the nicest properties is the existence of hubs. The definition for hubs is vague in the literature. In this paper we define a \textit{hub} in a network to a node whose degree is in the upper tail of the degree distribution. Intuitively, nodes with small degrees are usually connected through hubs. Therefore hubs in a network play an important role in information exchange and shortening the shortest paths between nodes. As we will discuss in section 3.1, scale-free networks have a smaller average geodesic distance than other networks. The existence of hubs is a significant difference between random networks and scale-free networks. In random networks, the expected degree is comparable for every node, and thus fewer hubs emerge. \\

The emergence of hubs can be explained by the growth algorithm of a scale-free network. A widely used model is the \textit{preferential attachment} model (\cite{albert2002statistical}):\\

The network begins with an initial connected network of $m_0$ nodes. New nodes are added to the network one at a time. Each new node is connected to $m \leq m_0$ existing nodes with a probability that is proportional to the number of edges that the existing nodes already have. Formally, the probability that the new node is connected to node $i$ is $\frac{k_i}{\sum_{j}k_j}$, where $k_i$ is the degree of the node $i$ and the sum is taken over all pre-existing nodes $j$. Numerical simulations (\cite{albert2002statistical}) indicated that this network evolves into a scale-free network with $\alpha=3$.\\

In Figure~\ref{tab:SF_ER} below, we illustrate the comparison between scale-free networks and Erd\H{o}s-R\'enyi random graph.\\

\begin{figure}[H]
 \begin{center}
 \includegraphics[scale=0.35]{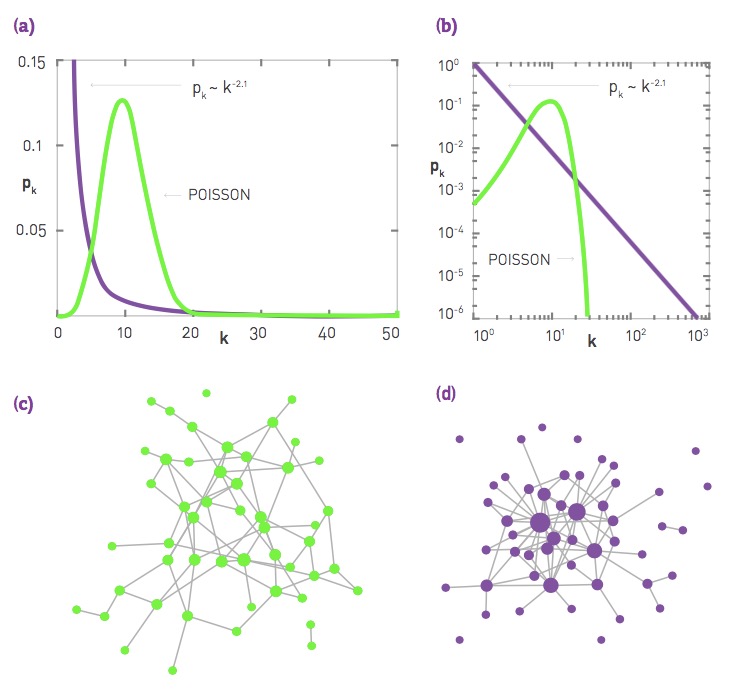}
 \caption{\textbf{Scale-free network vs. Erd\H{o}s-R\'enyi random graphs.} (\cite{barabasi2014})}
 \label{tab:SF_ER}
 \end{center}
\end{figure}

\begin{enumerate}[label=(\alph*)]
  \item Comparing a Poisson function with a power-law function ($\alpha=2.1$) on a linear plot. Both distributions have $<k>=11$. 
  \item The same curves as in (a), but shown on a log-log plot, allowing us to inspect the difference between the two functions in the high-$k$ regime. 
  \item An Erd\H{o}s-R\'enyi random network with $<k>=3$ and $n = 50$, illustrating that most nodes have comparable degree around $<k>$. The variation in degrees is very small.
  \item A scale-free network with $\alpha=2.1$ and $<k>=3$, illustrating that numerous small-degree nodes coexist with a few highly connected hubs. The size of each node is proportional to its degree, therefore the large ones are hubs in the network.
\end{enumerate}

\subsection{The Horvitz-Thompson Estimator and the Hansen-Hurwitz Estimator}

Suppose we have a population of elements $\{1, 2, ..., M\}$ and $y_i$ is the characteristic of interest associated with element $i$, $i=1, ..., M$. Let $t_y=\sum_{i=1}^{M}y_i$ denote the total of $y_i$'s. In order to estimate $t_y$ from samples of elements selected with unequal probabilities, we can use the Horvitz-Thompson estimator for samples drawn without replacement and the Hansen-Hurwitz estimator for samples drawn with replacement.\\

Suppose a sample of size $m$ is drawn without replacement from the population, and the inclusion probability for element $y_i$ is $\pi_i>0$. Let $Z_i$ be an indicator variable such that $Z_i=1$ if element $i$ is in the sample and 0 otherwise. The Horvitz-Thompson estimator (\cite{horvitz1952generalization}) of the population total $t_y$ is
\begin{equation}
\hat{t}_y^{HT} = \sum_{i=1}^{M}\frac{Z_i y_i}{\pi_i},
\end{equation}
with mean 
\begin{equation}
E(\hat{t}_y^{HT}) = t_y,
\end{equation}
and variance
\begin{equation}
Var(\hat{t}_y^{HT}) = \sum_{i=1}^{M}\sum_{k>1}^{M}(\pi_i\pi_j - \pi_ik)(\frac{y_i}{\pi_i}-\frac{y_k}{\pi_k})^2.
\end{equation}

Next suppose a sample of size $m$ is drawn with replacement in $m$ independent draws from the population,  and that on each draw the probability of selecting element $y_i$ is $\beta_i$. Let $Q_i$ denote the number of times element $y_i$ selected in the sample, so that $Q_1, ..., Q_M \sim \text{multinomial}(\beta_1, ..., \beta_M;m)$, $E(Q_i) = m\beta_i$, and $\sum_{i=1}^{N}Q_i=m$.The Hansen-Hurwitz estimator (\cite{hansen1943theory}) of the population total $t_y=\sum_{i=1}^{M}y_i$ is
\begin{equation}
\hat{t}_y^{HH} = \frac{1}{m}\sum_{i=1}^{M}\frac{Q_i y_i}{\beta_i},
\end{equation}
with mean
\begin{equation}
E(\hat{t}_y^{HH}) = t_y,
\end{equation}
and variance
\begin{equation}
Var(\hat{t}_y^{HH}) = \frac{1}{m}\sum_{i=1}^{M}\beta_i(y_i/\beta_i-t_y)^2.
\end{equation} 

More generally, we will consider sample selections that could be dependent with varying selection probabilities for different draws. Thus, we define a more general form of $\hat{t}_y^HH$ as
\begin{equation}
\hat{t_y}^{GHH} = \sum_{i=1}^{M}\frac{Q_i y_i}{E(Q_i)}.
\label{GHH}
\end{equation}
This is always unbiased for $t_y$ as long as $E(Q_i)>0$. The variance of $\hat{t}_y^{GHH}$ can be estimated if the sample is selected with replication.\\

Note that we can also estimate the total from a sample obtained by sampling with replacement by a Horvitz-Thompson estimator. If we reduce the sample obtained by sampling with replacement to a subsample by excluding the duplicates, we will get the subsample consisting of distinct elements from the population, which is analogous to a sample obtained by sampling without replacement but with random sample size. Therefore we can apply the idea of estimating the population total by Horvitz-Thompson estimator to the subsample, provided we can calculate $\pi_i$ terms.

\section{Related Work}

\subsection{The Small World Effect}
One of the most interesting and widely studied of network phenomena is \textit{the small world effect}: in many networks, the distances between nodes are surprisingly small. The first empirical study of this phenomenon goes back to Stanley Milgram's letter-passing experiment in the 1960s, in which he asked each of the randomly chosen “starter” individuals to try forwarding a letter to a designated “target” person living in the town of Sharon, MA, a suburb of Boston. It turned out that the letters made it to the target in a remarkably small number of steps, around six on average. Therefore, this phenomenon is also called ``six degrees of separation".\\

With complete network data and measuring methods available these days, it is possible to measure or estimate the distances between nodes, and the small world effect has been verified explicitly. In mathematical terms, the small-world effect is the condition that the mean distance $M$ is small. In fact, following the mathematical models, the mean distance for Erd\H{o}s-R\'enyi random graphs was shown to scale as $\log n$ (\cite{newman2010networks}, p.422). \\

What's more, analytical results have shown that the mean distances for scale-free networks are even smaller. \cite{chung2002average}, showed that for certain families of random graphs with given expected degrees the average distance is almost surely of order $\log n / \log \tilde{d}$ . Here $\tilde{d}$ denotes the second-order
average degree defined by $\tilde{d} = \frac{\sum w_i^2}{\sum w_i}$, where $w_i$ denotes the expected degree of the $i^{th}$ node. More specifically, for scale-free networks with $\alpha>3$, they proved that the average distance is almost surely of order  $\log n / \log \tilde{d}$. However, many Internet, social, and citation networks are scale-free networks with exponents in the range $2<\alpha<3$, for which the mean distance is almost surely of order $\log \log n$, but have diameter of order $\log n$ (subject some mild constraints for the average distance and maximum degree, see \cite{chung2002average} for details). This was followed by the study by \cite{cohen2003scale}, who showed, using analytical argument, that the mean distance $M \sim \log \log n$ for $2<\alpha<3$, $M \sim \log n /\log \log n$ for $\alpha=3$, and $M \sim \log n$ for $\alpha>3$.\\

To summarize, the small world effect on scale-free networks with $2 < \alpha < 3$ yields the nice property that the mean distance and the diameter are of scale $\log \log n$ and $\log n$ respectively. For instance, a scale-free network of size $n=10000$ has diameter only around $9$. A small diameter leads to a small range of SPL, and thus it's practical to estimate the SPLD, which consists of the percentage of dyads with a particular value of SPL for each possible value of SPL.

\subsection{Shortest Path Length Distribution}

The shortest paths  are of particular importance because they are likely to provide the fastest and strongest interaction between nodes in a network (\cite{katzav2015analytical}). Up to now, measures such as the diameter and the mean distance have been studied extensively, but the entire shortest path length distribution (SPLD) has apparently attracted little attention. This distribution is of great importance as it's closely related to dynamic properties such as velocities of network spreading processes (\cite{bauckhage2013weibull}). More specifically, it plays a key role in the temporal evolution of dynamical processes on networks, such as signal propagation, navigation, and epidemic spreading (\cite{pastor2001epidemic}). \\

\cite{katzav2015analytical} showed two complementary analytical approaches for calculating the distribution of shortest path lengths in Erd\H{o}s-R\'enyi networks, based on recursion equations for the shells around a reference node and for the paths originating from it. However, Erd\H{o}s-R\'enyi graphs are not widely observed in real networks and are often only of research interest because of their simple structure. In practice, we are more interested in a wider class of networks.\\

Other researchers such as \cite{bauckhage2013weibull} have characterized shortest path histograms of networks by the Weibull distributions. Empirical tests with different graph topologies, including scale-free networks, have confirmed their theoretical prediction. However, each real network has its own parameter values of the Weibull distribution, and it is hard to find those values without full access to the network. What's more, even if we can measure the shortest distance between any pair of nodes in a network, it is very time-consuming when the network is large (\cite{potamias2009fast}). Therefore, here in this paper, we consider estimating the SPLD of a population graph by the sample data generated by random walks.

\subsection{Ability of Random Walks to Recover Shortest Paths}
The strong ability of random walks to discover the shortest paths in networks with large degree variability was shown by \cite{ribeiro2012multiple}. They found that the ability of random walks to find shortest paths bears no relation to the paths they take, but instead relies on the large variance of the degree distribution of the network. \\

They proved two important results for networks with large degree variability. First, even with a relatively small number of steps, a single random walk is able to traverse a large fraction of edges. Let $<k^r>$ denote the $r^{th}$ moment of the degree distribution. They show that for a single random walk with $t$ steps, the number of edges discovered by the random walk is approximately $\frac{<k^2>-<k>}{<k>}t$, which is very large for networks with large variance in degree distribution. Second, two random walks cross with high probability after a small percentage of nodes have been visited. The first result indicates that the observed SPLs in the induced subgraph are very likely to be the true SPLs in the population. With a large fraction of edges visited by the random walk, the true shortest paths are very likely to be observed. The second result implies that a single random walk has the potential the explore a large area in the population network, instead of staying around the small area close to itself. This property provides the possibility of using a single random walk to uncover the true SPLs. We will verify this property in section 5.3. These observations provide the possibility of using random walks to uncover shortest paths in networks with large degree variability. \\

Their simulation results on some real networks are also very promising. For most real-world networks they tested, more than $65 \%$ of the shortest paths observed in the sampled graph by random walk sampling are the true shortest paths in the parent graph, and more than $90 \%$ of the shortest paths observed in the sampled graph by random walk sampling are within one hop of the true shortest paths in the parent graph. The only exception is a network whose degree variability measured by $\frac{<k^2>-<k>}{<k>}t$ is much smaller than other networks.

\subsection{Estimating Shortest Distances by Landmarks}
Computing the shortest distance, i.e., the length of the shortest path between arbitrary pairs of nodes, has been a prominent problem in computer science. In an unweighted graph with $n$ nodes and $m$ edges, the shortest distances between one node and all other nodes can be computed by the Breadth First Search (BFS) algorithm in time $O(m+n)$ (\cite{potamias2009fast}). To measure the distances between all pairs of nodes, one can implement the BFS algorithm $n$ times in time $O(n^2 + mn)$, which is quadratic in the number of nodes. Therefore, in large networks, computing the exact shortest distances between all pairs of nodes is computationally expensive. To improve the efficiency, several fast approximation algorithms have been developed recently.\\

Most of the approximation algorithms are landmark-based methods. They start from selecting a small set of nodes called landmarks. Then the actual distances from each landmark to all other nodes in the graph are computed by BFS and stored in memory. By using the precomputed shortest distances from the landmarks, the distance between an arbitrary pair of nodes can be computed in almost constant time. The algorithm proposed by \cite{potamias2009fast} is one of the landmark-based methods to quickly estimate the the length of the point to point shortest path.\\

Their algorithm is based on the triangle inequalities for the geodesic distance. That is, given any three nodes $s$, $u$, and $t$, the geodesic distances between them satisfy the following inequalities:
\begin{equation}
l_{st} \leq l_{su} + l_{ut},
\label{upper}
\end{equation}
\begin{equation}
l_{st} \geq |l_{su} - l_{ut}|.
\end{equation}
Note that if $u$ lies on one of the shortest paths from $s$ to $t$, then inequality (\ref{upper}) holds with equality.\\

In the pre-computing step, a set of $d$ landmarks $D$ are selected from the graph, and the actual distances between each landmark and all other nodes are computed by BFS. In the estimating step, by the above inequalities, the actual geodesic distance between node $s$ and $t$ satisfies:
\begin{equation}
L \leq l_{st} \leq U,
\end{equation}
where
\begin{equation}
L = max_{j \in D}|l_{sj} - l_{jt}|,
\end{equation}
\begin{equation}
U = min_{i \in D}\{l_{si} + l_{it}\}.
\end{equation}

By experiments, \cite{potamias2009fast} proposed simply using the upper bound $U$ as an estimate to the geodesic distance. That is,
\begin{equation}
l_{st} \approx min_{i \in D}\{l_{si} + l_{it}\}.
\end{equation}
This algorithm takes $O(d)$ time to approximate the distance between a pair of nodes and requires $O(dm+dn)$ space for the pre-computation data.\\

Note that the approximation will be very precise if many shortest paths pass through the landmarks. That is, the best set of landmarks consists of the most "central" nodes in the graph, and more specifically, the nodes with high betweenness centralities. In graph $G$, let $n_{st}^i$ be the number of shortest paths between node $s$ and node $t$ passing node $i$, and $g_{st}$ be the total number of shortest paths between $s$ and $t$, the \textit{betweenness centrality} of node $i$ is defined to be $\sum_{st}\frac{n_{st}^i}{g_{st}}$. Intuitively, it measures the fraction of shortest paths passing node $i$. Generally, nodes with high degrees usually have high betweenness centralities but nodes with high betweenness centralities don't always have high degrees. One example would be a graph consisting of two clusters which are connected trough a single node. The connecting node has only degree $2$ but its betweenness centrality is really high.\\

Measuring the betweenness centrality of a node requires the information of shortest paths between all nodes in the sample, which can not be observed from the sample. As an alternative, \cite{potamias2009fast} came up with two basic strategies based on other centrality measures for selecting landmarks: (i) high degree nodes and (ii) nodes with high estimated \textit{closeness centrality}, where the closeness centrality is the inverse of the average distance from a node to all other nodes. They defined the estimation error to be the average of $|\hat{l}-l|/l$ across all pairs of sampled nodes, where $l$ is the actual distances and $\hat{l}$ is the approximation. Regarding to the size of the set of landmarks, they found from the application to some real networks that, with $100$ landmarks, the estimation error is at less than $10\%$ in $3$ of the $5$ real networks, and between $10\%$ and $20\%$ in the other 2 real networks.

\section{Proposed Method}

\subsection{Intuition}
Recall that in a scale-free network, most nodes with small degrees are connected through hubs. Our approach is based on the following intuition: random walks in scale-free networks usually take steps along the shortest paths between pairs of nodes. This nice behavior is attributed to the existence of hubs. \\

Consider an extreme case of a network with only one hub to which all other nodes are connected. Then the random walk always goes back to the hub before moving to another node, which indeed is following the shortest path of length 2 between the nodes before and after the hub. Next consider a network with multiple hubs, but still, all other nodes are connected only to the hubs. In this case a random walk starting from any node will have to go back to the hub to which the node is connected to get to another node, which forces the random walk to travel along the shortest path for a pair of nodes. \\

\begin{figure}[H]
 \begin{center}
 \includegraphics[scale=0.3]{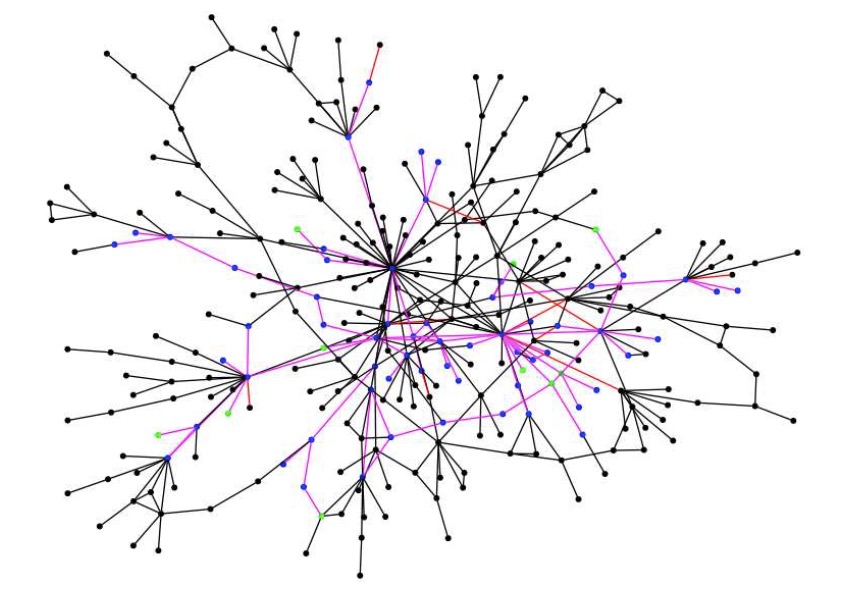}
 \caption{\textbf{Illustration of a RW sample path.} The green nodes are the starting nodes, the blue nodes are nodes visited by the random walkers, and the purple edges are the edges used by the walkers to explore the graph.(\cite{ribeiro2012multiple})}
 \label{RW}
 \end{center} 
\end{figure}

More generally, if there are some but very few connections between nodes which are not hubs, a random walk might have the chance to traverse a path that is not the shortest path between two nodes, but the chance is small. Figure \ref{RW} shows how multiple random walks recover shortest paths in a scale-free network.

\subsection{Problem Definition}
Consider a connected and undirected network $G=(V,E)$ with $n$ nodes, $m$ edges, and diameter $L$. Then the shortest path length distribution (SPLD) of $G$ is defined as
\begin{equation}
f_l = \frac{N_l}{N}, l=1,...,L
\end{equation}
where $N_l$ is the number of dyads with SPL $l$, and $N=\sum_{l=1}^{L}N_l={}_nC_2$ is the total number of dyads (pairs of nodes) in $G$.

\subsection{Sampling Algorithm}
For a given network $G=(V,E)$, we first take a simple random sample of $H$ distinct nodes $U=\{u_1,..., u_H\}$, and start a random walk from each of them. The $H$ random walks are independent after the starting nodes. We define the sampling budget, denoted by $\beta, \text{ } 0<\beta<1$, to be the ratio of total steps of the $H$ random walks to the networks size $n$, and let each random walk take $B=\beta n/H$ steps. \\

Let $X(h)=(X_1^{(h)},...,X_B^{(h)}), \text{ } h=1,...,H$, denote the sequence of nodes visited by the $h^{th}$ walker. Let $V(h)$  denote the set of distinct nodes visited by the $h^{th}$ walker, and $|V(h)|$ denote the number of nodes in set $V(h)$. Note that $|V(h)| \leq B$ as a node can be revisited during the random walk. Let $E(h)$ denote the set of edges in $E$ that have both endpoints in $V(h)$.\\

Let $V^*=\bigcup_{h=1}^{h=H}V(h)$ denote the set of distinct nodes visited by the any of the $H$ random walks, and $E^*$ denote the set of edges in $E$ that have both of their endpoints in $V^*$. Then $G^*=(V^*, E^*)$ is the induced subgraph obtained by connecting nodes in $V^*$ using edges in $E^*$. The observed shortest path length between any two sampled nodes will be measured from $G^*$.

\subsection{Estimating Method}

In order to estimate the fraction $f_l$ of dyads with SPL $l$, we need to first estimate $N_l$, the number of dyads with SPL $l$ in the population graph. Let $\hat{N}_l$ denote the estimate for $N_l$, and $f_l$ can be estimated by $\hat{f}_l = \frac{\hat{N}_l}{N}$. Note that sometimes we want to use a ratio estimator $\hat{f}_l^r = \frac{\hat{N}_l}{\hat{N}}$, in which case we also estimate $N$, the total number of dyads in the population graph.\\

\subsubsection{The Unweighted Estimator}
A naive approach to estimate population SPLD is to simply use the SPLD of the induced subgraph $G^*$ as an estimate. Let $N_l^*$ denote the number of dyads with SPL $l$ in $G^*$, and $N^*$ denote the total number of dyads in $G^*$, the unweighted estimator for $f_l$ is
\begin{equation}
\hat{f}_l^{uw} = \frac{N_l^*}{N^*}, \text{ } l=1,...,L.
\label{UW_Nl}
\end{equation}

However, this simple estimator may suffer from two sources of bias. First, the dyads are sampled with unequal probabilities due to the nature of random walk sampling. More specifically, dyads with shorter SPLs are more likely to be sampled than those with longer SPLs. Therefore, with the unweighted estimator, $f_l$ for small value of $l$ is likely to be over estimated, and $f_l$ for large value of $l$ is likely to be under estimated. Second, the observed SPL in $G^*$ might be longer than the actual SPL $G$, and thus $f_l$ for small value of $l$ is likely to be under estimated, and $f_l$ for large value of $l$ is likely to be over estimated. \\

As discussed in Section 3.3 and 4.1, the bias from not observing the actual SPL is negligible in networks with large degree variability. We will discuss this issue in details in section 4.4.4.  In the section 4.4.2 and section 4.4.3, we will develop estimators that deal with the unequal selection probabilities of dyads.

\subsubsection {The Hansen-Hurwitz Estimator}
Let $s = \{X(1), X(2), ..., X(H)\}$ denote the set of sequences of nodes visited by $H$ random walks, including duplicates, and let $|s|=H\cdot B$ denote the size of $s$. Let $I(X^{(h)}_b=i)$ denote an indicator variable taking the value 1 if node $i$ is visited at $b^{th}$ step in $h^{th}$ random walk, and zero otherwise. Let $q_i=\sum_{h=1}^{H}\sum_{b=1}^{B}I(X^{(h)}_b=i), \text{ } i=1,...,n$ denote the number of times node $i$ appears in sample $s$, and define $\phi_i=E(q_i)/|s|$. 
We assume  $0< E(q_i)< |s|$ $\forall i$, and thus $0 < \phi_i < 1$ $\forall i$. Since $\sum_{i=1}^{n}q_i=|s|$, $\sum_{i=1}^{n}\phi_i=1$. Therefore, the $\phi_i$'s form a probability distribution over the $n$ nodes.\\

Let $r, \text{ } r=1,...,N$ represent dyad $(i,j), \text{ } i=1,...,n-1, \text{ } j=i+1,...,n$ in the population graph. Let $S = \{(X_{b_1}^{(h_1)}, X_{b_2}^{(h_2)}):h_1, h_2 \in \{1,...,H\}, b_1, b_2 \in \{1,...,B\}, X_{b_1}^{(h_1)}\neq X_{b_2}^{(h_2)}\}$ denote the set of dyads whose members are any two distinct nodes in $s$. That os, $S$ is the sequence of dyads visited by the $H$ random walks, including duplicates. Define $Q_r=q_iq_j, \text{ } i=1,...,n-1, \text{ } j=i+1,...,n$ as the number of times dyad $r$ appears in sample $S$, and let $|S|=\sum_{r=1}^{N}Q_r$ denote the size of $S$. Notice that there may be duplicates in the sample of nodes $s$, but to a dyad, we only include pairs consisting of  two different nodes, therefore $|S|$ is a random variable with $|S| = \binom{|s|}{2}-\sum_{i=1}^{n}\binom{q_i}{2}$. Define $\psi_r=\frac{E(Q_r)}{E(|S|)}$ and assume $0<E(Q_r)<|S|$ $\forall r$, therefore $0 < \psi_r < 1$ $\forall r$. Since $\sum_{r=1}^{N}Q_r=|S|$, $\sum_{r=1}^{N}\psi_r=1$. Therefore, the $\psi_r$'s form a probability distribution over the $N$ dyads. \\

Let $l_r\in\{1,...,L\}$ denote the true SPL of dyad $r$ in the population graph. Let $y_r^l$, $r \in \{1, ..., N\}$ and $l \in \{1, ..., L\}$, denote an indicator variable taking value $y_r^l=1$ if $l_r=l$ and zero otherwise. Thus $N_l=\sum_{r=1}^{N}y_r^l$ is the number of dyads with SPL $l$ in the population, and $N=\sum_{l=1}^{L}\sum_{r=1}^{N}y_r^l$ is the total number of dyads in the population.\\

 According to \ref{GHH}, the generalized Hansen-Hurwitz estimator for $N_l$ is
\begin{equation}
\hat{N}_l^{GHH} = \frac{1}{|S|}\sum_{r=1}^{N}\frac{Q_ry_r^l}{\psi_r}, \text{ } l=1,...,L
\label{HH_Nl}
\end{equation}

The generalized Hansen-Hurwitz estimator for $N$ is
\begin{equation}
\hat{N}^{GHH} = \frac{1}{|S|}\sum_{r=1}^{N}\frac{Q_r}{\psi_r}, \text{ } l=1,...,L
\label{HH_N}
\end{equation}

In order to apply (\ref{HH_Nl}) and (\ref{HH_N}) we need to compute or estimate $\psi_r$. We first recall some definitions and results for Markov chains. We call a sequence of random variables $\{X_t: t=1, 2, ...\}$ a \textit{discrete-time Markov chain (DTMC)} if it satisfies 
\begin{equation}
P(X_{t+1}=i_{t+1}|X_t=i_t, X_{t-1}=i_{t-1}, .., X_1=i_1)=P(X_{t+1}=i_{t+1}|X_t=i_t),
\end{equation}
for all $t\geq1$ and $i_1, i_2, ..., i_{t+1} \in \Omega$, where $\Omega$ is a finite or countable state space. \\

A DTMC is \textit{finite} if $\Omega$ is finite. A DTMC is \textit{homogeneous} if it satisfies 
\begin{equation}
P(X_{t+1}=j|X_t=i)=P_{i,j} \text{ for all } i, j\in \Omega, \text{ independent of } t.
\end{equation}
We call the probabilities $P_{i, j}$'s the \textit{transition probabilities}. Let $\boldsymbol{P}$ denote a matrix with element $P_{i, j}$ at its position of $i^{th}$ row and $j^{th}$ column. We call $\boldsymbol{P}$ the \textit{transition matrix} for a homogeneous DTMC. Since we will only consider finite DTMCs in this paper, we denote $\Omega = \{1, 2, ..., n\}$ for simplicity. \\

Let $p_i(t)$ denote the probability that $\{X_t\}$ is in state $i$ at time $t$, and let $\boldsymbol{p}(t)=(p_1(t), p_2(t), ..., p_n(t))^T$ denote the vector of probabilities. For a finite homogeneous DTMC we have
\begin{equation}
\boldsymbol{p}^T(t+1)= \boldsymbol{p}^T(t)\boldsymbol{P}.
\end{equation}

A probability vector $\boldsymbol{p}=(p_1, p_2, ..., p_n)^T$ is called a \textit{stationary distribution} for a homogeneous DTMC with transition matrix $\boldsymbol{P}$, if it satisfies 
\begin{equation}
\boldsymbol{p}^T=\boldsymbol{p}^T\boldsymbol{P}.
\end{equation}

State $j$ is said to be \textit{accessible} from state $i$ if $P^n_{i,j}>0$ for some $n\geq 0$. If state $i$ is accessible from state $j$ and state $j$ is accessible from state $i$, $i$ and $j$ are said to \textit{communicate}. A DTMC is called \textit{irreducible} if all of its states communicate with each other. A state $i$ is \textit{aperiodic} if the greatest common divisor of $\{n\geq0: P_{i,i}^n>0\}$ is $1$. A DTMC is called \textit{aperiodic} if all of its states are aperiodic.\\

\begin{itemize}

\item \textbf{Proposition 1:} (\cite{newman2010networks} p.157-159) A single random walk $\{X_t\}$ on a graph $G=(V,E)$ of size $n$ is a finite homogeneous DTMC with a stationary distribution $\boldsymbol{p}=(\frac{k_1}{K}, ..., \frac{k_n}{K})^T$, where $K=\sum_{w}k_w$.

\item \textbf{Proof:} Consider a random walk $\{X_t\}$ that starts at a certain node and takes $t$ steps. Suppose $\{X_t\}$ is at node $i$ at time $t-1$, then the probability that it will be at node $j\neq i$ at time $t$ is $1/k_i$, by the definition of random walk sampling in section 2.2, given that $i$ is connected to $j$, i.e., $A_{ij}=1$. That is
\begin{equation}
P(X_t=j|X_{t-1}=i)=\frac{A_{ij}}{k_i}.
\end{equation}

Therefore, $\{X_t\}$ is a homogeneous DTMC with finite state space $\{1, 2, ..., n\}$ and transition probabilities $P_{i, j}=\frac{A_{ij}}{k_i}$. Let $\boldsymbol{P}$ denote the transition matrix of $\{X_t\}$, then $\boldsymbol{P}=\boldsymbol{D}^{-1}\boldsymbol{A}$, where $\boldsymbol{D}$ is the diagonal matrix with elements $k_i$'s for $i=1,...,n$. \\

Let $\boldsymbol{p} = (\frac{k_1}{K}, \frac{k_2}{K}, ..., \frac{k_n}{K})^T$, where $K=\sum_{w}k_w$. 
\begin{align}
\boldsymbol{p}^T\boldsymbol{D}^{-1}\boldsymbol{A}
& = \begin{pmatrix}
  \frac{k_1}{K} & \frac{k_2}{K} & ... & \frac{k_n}{K}
\end{pmatrix}
\begin{pmatrix}
  \frac{A_{11}}{k_1} & \frac{A_{12}}{k_1} & ... & \frac{A_{1n}}{k_1} \\ 
  \frac{A_{21}}{k_2} & \frac{A_{22}}{k_2} & ... & \frac{A_{2n}}{k_2} \\
  \vdots & \vdots & \ddots & \vdots \\
  \frac{A_{n1}}{k_n} & \frac{A_{n2}}{k_n} & ... & \frac{A_{nn}}{k_n}\\
\end{pmatrix}\\
& = \begin{pmatrix}
  \sum_{i=1}^{n}\frac{k_i}{K}\frac{A_{i1}}{k_i}
  & \sum_{i=1}^{n}\frac{k_i}{K}\frac{A_{i2}}{k_i}
  & ...
  & \sum_{i=1}^{n}\frac{k_i}{K}\frac{A_{in}}{k_i}
\end{pmatrix}\\
& = \begin{pmatrix}
  \frac{1}{K}\sum_{i=1}^{n}A_{i1}
  & \frac{1}{K}\sum_{i=1}^{n}A_{i2}
  & ...
  & \frac{1}{K}\sum_{i=1}^{n}A_{in}
\end{pmatrix}
&=
\begin{pmatrix}
  \frac{k_1}{K}
  & \frac{k_2}{K}
  & ...
  & \frac{k_n}{K}
\end{pmatrix}
= \boldsymbol{p}^T
\end{align}

That is, $\boldsymbol{p}^T = \boldsymbol{p}^T\boldsymbol{P}$. Since $p_i>0$ and $\sum_{j}p_j=1$, $\boldsymbol{p}$ is a stationary distribution for $\{X_t\}$. \\

\item \textbf{Proposition 2:} If $G$ is connected and has at least one triangle, the finite homogeneous DTMC $\{X_t\}$ from \textbf{Proposition 1}  is irreducible and aperiodic. 

\item \textbf{Proof:} Since $G$ is connected, any node in the is accessible by any other node. That is, all states of $\{X_t\}$ communicate with other, and thus $\{X_t\}$ is irreducible. For any node in $G$, it can be either in a triangle or not. Suppose $i$ is any node in a triangle, then starting from itself, $i$ can be reached by either 2 steps or 3 steps, that is $P_{i,i}^2>0$ and $P_{i,i}^3>0$. Therefore $i$ is an aperiodic state. Consider any node $j$ which is not in a triangle and suppose that its shortest distance to node $i$ is $l$, then starting from itself, $j$ can be reached by either $2l+2$ steps or $2l+3$ steps, that is $P_{i,i}^{2l+2}>0$ and $P_{i,i}^{2l+3}>0$. Therefore $j$ is also an aperiodic state. Since all states in $\{X_t\}$ are aperiodic, $\{X_t\}$ is aperiodic.\\

\item \textbf{Proposition 3:} If a single random walk $\{X_t\}$ initiates from its stationary distribution $\boldsymbol{p}$ on a connected graph $G$ with at least one triangle, then $\phi_i = E(q_i)/t = k_i/K$, and $lim_{t \rightarrow \infty}\psi_r=\alpha k_ik_j$, where $\alpha=2[(\sum_{w}k_w)^2-\sum_{w}k_w^2]^{-1}$, and $K = \sum_{w=1}^{n}k_w$.

\item \textbf{Proof:} Let $\boldsymbol{q} = (q_1, q_2, ..., q_n)^T$, where $q_i$ = number of times node $i$ appears in the sample, and $\boldsymbol{p} = (p_1, p_2, ..., p_n)^T$, where $p_i$ = $\frac{k_i}{K}$. According to Anderson's (1989) results for irreducible and aperiodic Markov chains, 
\begin{equation}
E(\boldsymbol{q})=\boldsymbol{p}t,
\label{Anderson1}
\end{equation}
and
\begin{equation}
\lim_{t\rightarrow\infty}\frac{Cov(\boldsymbol{q})}{t}= C,
\label{Anderson2}
\end{equation}
where $C$ is a square matrix with constant elements.\\

From (\ref{Anderson1}), we have $\frac{E(q_i)}{t} = \frac{k_i}{K}$, for $i=1, ..., n$.\\

In general $a_n=O(b_n)$ indicates $\lim_{t \rightarrow \infty}a_n/b_n=c$, where $c$ is a constant, and $a_n=o(b_n)$ indicates $\lim_{t \rightarrow \infty}a_n/b_n=0$, so we have
\begin{equation}
Cov(q_i, q_j)=o(t^2), \text{ and } Var(q_i)=o(t^2) \text{ } \forall i.
\end{equation}

The expected number of times dyad $r$ appears in sample $S$ is
\begin{align}
E(Q_r) = E(q_iq_j) = E(q_i)E(q_j) + Cov(q_i, q_j) = p_ip_jt^2+o(t^2)
\end{align}

The expected number of dyads (including duplicates) in sample $S$ is
\begin{align}
E(|S|) &= \binom{t}{2}-\sum_{i=1}^{n}E(\frac{q_i(q_i-1)}{2})\\
&=\binom{t}{2}-\frac{1}{2}\sum_{i=1}^{n}(E(q_i^2)-E(q_i)) \\
&=\binom{t}{2}-\frac{1}{2}\sum_{i=1}^{n}(E^2(q_i)-E(q_i)+Var(q_i)) \\
&=\binom{t}{2}-\frac{1}{2}\sum_{i=1}^{n}tp_i(tp_i-1)+o(t^2)\\
&= \frac{1}{2}t(t-1) -\frac{1}{2}(t^2\sum_{i=1}^{n}p_i^2-t) + o(t^2)\\
& = \frac{1}{2}(1-\sum_{i=1}^{n}p_i^2)t^2 + o(t^2) 
\end{align}

In the long run, the expected fraction that dyad $r$ appears in sample $S$ is
\begin{align}
\lim_{t\rightarrow\infty}\psi_r &= \lim_{t\rightarrow\infty} \frac{E(Q_r)}{E|S|}\\
& = \lim_{t\rightarrow\infty} \frac{2p_ip_jt^2 + o(t^2)}{(1-\sum_{i=1}^{n}p_i^2)t^2+o(t^2)}\\
& = \frac{2p_ip_j}{1-\sum_{i=1}^{n}p_i^2}\\
&=\frac{2\frac{k_ik_j}{(\sum_{w}k_w)^2}}{1-\frac{\sum_{w}k_w^2}{(\sum_{w}k_w)^2}}\\
&=\frac{2k_ik_j}{(\sum_{w}k_w)^2-\sum_{w}k_w^2} 
\end{align} 

For simplicity we can write $lim_{t \rightarrow \infty}\psi_r=\alpha k_ik_j$, where $\alpha=2[(\sum_{w}k_w)^2-\sum_{w}k_w^2]^{-1}$.\\

\end{itemize}

Therefore, the generalized Hansen-Hurwitz estimator for $N_l$ is
\begin{equation}
\hat{N}_l^{GHH} =\frac{1}{|S|}\sum_{r=1}^{N}\frac{Q_ry_r^l}{\alpha k_i k_j}, \text{ } l=1,...,L, 
\end{equation}

and the generalized Hansen-Hurwitz estimator for $N$ is
\begin{equation}
\hat{N}^{GHH} = \frac{1}{|S|}\sum_{r=1}^{N}\frac{Q_r}{\alpha k_i k_j}, \text{ } l=1,...,L.
\end{equation}

The generalized Hansen-Hurwitz estimator for the fraction of dyads with SPL $l$ is
\begin{equation}
\hat{f}_l^{GHH} = \frac{\hat{N}_l^{HH}}{N}=\frac{\sum_{r=1}^{N}\frac{Q_ry_r^l}{\alpha k_ik_j}}{|S|N}, \text{ } l=1,...,L,
\end{equation}

and the generalized Hansen-Hurwitz ratio estimator for the fraction of dyads with SPL $l$ is
\begin{equation}
\hat{f}_l^{GHH.r} = \frac{\hat{N}_l^{HH}}{\hat{N}^{HH}}=\frac{\sum_{r=1}^{N}\frac{Q_ry_r^l}{k_ik_j}}{\sum_{r=1}^{N}\frac{Q_r}{k_ik_j}}, \text{ } l=1,...,L
\label{HH.ratio}
\end{equation}

\subsubsection {The Horvitz-Thompson Estimator}
In the Hansen-Hurwitz estimator illustrated above, we take the average of all observed dyads, including duplicates, to estimate $N_l$ and $N$. Alternatively, we can consider applying the Horvitz-Thompson estimator to the subsample obtained by excluding duplicate observations.\\

Let $s^* = V^*$ denote set of distinct nodes visited by $H$ random walks, and $|s^*| = \sum_{h=1}^{H}|V(h)|$ denote the sample size of $s^*$. Since $s^*$ is derived from $s$ by excluding the duplicates, $|s^*|$ is a random variable depending on $s$. Let $z_i, \text{ } i=1,...,n$ denote the number of times node $i$ appears in sample $s^*$. In our case $z_i$ is an indicator variable such that $z_i=1$ if $i\in s^*$ and zero otherwise. Let $\tau_i=E(z_i)$ denote the inclusion probability of node $i$ in the subsample $s^*$, which is indeed the probability that node $i$ ever appears in sample $s$.
Since $\sum_{i=1}^{n}z_i=|s^*|$, we have $\sum_{i=1}^{n}\tau_i=E(|s^*|)$.\\

Let $S^*$ denote the set of all pairs of nodes in $s^*$, and let $|S^*|$ denote the size of $S^*$. Let $Z_r, \text{ } i=1,...,n-1, \text{ } j=i+1,...,n$ denote the number of times dyad $r=(i,j)$ appears in sample $S^*$. In our case $Z_r$ is an indicator variable such that $Z_r=1$ if $r\in S^*$ and zero otherwise. Let $\pi_r=E(Z_r)$ denote the inclusion probability of dyad $r$ in the subsample $S^*$, which is indeed the probability that dyad $r$ ever appears in sample $S$.
Since $\sum_{r=1}^{N}Z_r=|S^*|$, we have $\sum_{r=1}^{N}\pi_r=E(|S^*|)$.\\

Due to the lack of knowledge about the full network $G=(V,E)$ as well as computational considerations, we will use an approximation for estimating $\pi_r$, $r \in S^*$. If a single random walk $\{X_t\}$ initiates from its stationary distribution $\boldsymbol{p}$ on a connected graph $G$ with at least one triangle, in the long run,
\begin{equation}
\pi_r \approx \tau_i\tau_j, \text{ for } r=1, 2, ..., N,
\end{equation}
where
\begin{eqnarray}
\tau_i = \frac{|s^*|}{\sum_{i=1}^{n}\theta_i}\theta_i \text{ for } i=1, 2, ...,n,
\label{tau}
\end{eqnarray}
and 
\begin{equation}
\theta_i=1-(1-\frac{k_i}{\sum_{w}k_w})^t \text{ for } i=1, 2, ...,n.
\end{equation}

\begin{itemize}
\item \textbf{Heuristic proof:} To derive the expected number of appearances of dyads in $S$, we used (\ref{Anderson2}) but did not need to use the form of the matrix $C$. A simple sampling model that satisfies (\ref{Anderson1}) and (\ref{Anderson2}) is multinomial sampling with $t$ draws and probability $p_i=\frac{k_i}{\sum_{w}k_w}$ for node $i$ to be sampled at each draw. For multinomial sampling, 
\begin{align}
E(q_i) = tp_i, 
\end{align}
and
\begin{align}
Cov(q_i, q_j)=\left\{
                \begin{array}{ll}
                  -tp_ip_j \text{, } i \neq j,\\
                  tp_i(1-p_i) \text{, } i=j,
                \end{array}
                \right.
\end{align}
and hence (\ref{Anderson1}) and (\ref{Anderson2}) are satisfied.

Under multinomial sampling, the probability that node $i$ is ever included in the sample by step $t$ is
\begin{equation}
\theta_i = 1-(1-p_i)^t.
\end{equation}

The joint probability that $i$ and $j$ are both included in the sample is 
\begin{equation}
\theta_r = \theta_{ij} = \sum_{x=1}^{t-1}P(z_i=1|q_j=x)P(q_j=x).
\end{equation}
Note that 
\begin{equation}
P(q_j=x) = \binom{t}{x}p_j^x(1-p_j)^{t-x},
\end{equation}
and
\begin{equation}
P(z_i=1|q_j=x) = 1-(1-\frac{p_i}{1-p_j})^{t-x},
\end{equation}
so
\begin{align}
\theta_r& = \sum_{x=1}^{t-1}\binom{t}{x}p_j^x(1-p_j)^{t-x}
[1-(1-\frac{p_i}{1-p_j})^{t-x}]\\
&= \sum_{x=1}^{t-1}\binom{t}{x}p_j^x(1-p_j)^{t-x}
-\sum_{x=1}^{t-1}\binom{t}{x}p_j^x(1-p_i-p_j)^{t-x}\\
&= 1-(1-p_ij)^t-p_j^t-[(1-p_i)^t-(1-p_i-p_j)^t-p_j^t]\\
&= 1-(1-p_i)^t-(1-p_j)^t+(1-p_i-p_j)^t.
\end{align}

Since
\begin{align}
\theta_i\theta_j&=[1-(1-p_i)^t][1-(1-p_j)^t]\\
& = 1-(1-p_i)^t-(1-p_j)^t+(1-p_i-p_j+p_ip_j)^t\\
& \approx 1-(1-p_i)^t-(1-p_j)^t+(1-p_i-p_j)^t \text{ if } p_ip_j \text{ is negligible},
\end{align}
and as $p_ip_j$ is verified to be negligible by simulations in this case, we can estimate $\theta_r$ by
\begin{equation}
\theta_r \approx \theta_i\theta_j.
\end{equation}

The only problem in approximation by multinomial sampling is that we assume the draws are independent, while it is not the case in random walk sampling since a node can't be sampled twice consecutively. Therefore, $\theta_i$ under the multinomial sampling model over estimates $\tau_i$, the inclusion probability of node $i$ in random walk sampling. To adjust for the overestimation, we can use the one of the following two approaches to estimate $\tau_i$, and then estimate $\pi_r$ by
\begin{equation}
\pi_r \approx \tau_i \tau_j.
\end{equation}

\textbf{Approach 1:} Using the fact $\sum_{r=1}^{n}\tau_i=E(|s^*|)$ as a constraint for $\tau_i$, we can estimate $\tau_i$ by 
\begin{equation}
\tau_i = \frac{|s^*|}{\sum_{i=1}^{n}\theta_i}\theta_i,
\label{HT1}
\end{equation}

\textbf{Approach 2:} Using the fact $\sum_{i \in s^*}\tau_i^{-1}=n$, we can choose the exponent $t^*<t$ for the random walking sampling such that 
\begin{equation}
(\sum_{i \in s^*}\frac{1}{1-(1-\phi_i)^{t^*}}-n)^2
\end{equation}
is minimized, and estimate $\tau_i$ by 
\begin{equation}
\tau_i = 1-(1-\phi_i)^{t^*}.
\label{HT2}
\end{equation}

Simulation results have shown that both (\ref{HT1}) and (\ref{HT2}) can provide a good estimation for $\tau_i$. 

\end{itemize}

The Horvitz-Thompson estimator for $N_l$ is 
\begin{equation}
\hat{N}_l^{HT} = \sum_{r=1}^{N}\frac{Z_ry_r^l}{\pi_r}, \text{ } l=1,...,L,
\end{equation}

 and the Horvitz-Thompson estimator for $N$ is 
\begin{equation}
\hat{N}^{HT} = \sum_{r=1}^{N}\frac{Z_r}{\pi_r}
\end{equation}

The Horvitz-Thompson estimator for the fraction of dyads with SPL $l$ is
\begin{equation}
\hat{f}_l^{HT} = \frac{\hat{N}_l^{HT}}{N}=\frac{\sum_{r=1}^{N}\frac{Z_ry_r^l}{\pi_r}}{N}, \text{ } l=1,...,L
\end{equation}

and the Horvitz-Thompson ratio estimator for the fraction of dyads with SPL $l$ is
\begin{equation}
\hat{f}_l^{HT.r} = \frac{\hat{N}_l^{HT}}{\hat{N}^{HT}}=\frac{\sum_{r=1}^{N}\frac{Z_ry_r^l}{\pi_r}}{\sum_{r=1}^{N}\frac{Z_r}{\pi_r}}, \text{ } l=1,...,L
\end{equation}

\subsubsection{Approximating actual SPLs between sampled nodes}
As discussed in section 3.4, in a network with $n$ nodes and $m$ edges, the time complexity to measure the actual distances between all pairs of nodes is $O(mn+n^2)$. This is computationally expensive for large networks. With our proposed estimators discussed above, we only need measure  the distances between sampled nodes to estimate the SPLD of the population graph. Let $\beta^*$ denote the fraction of nodes in the induced subgraph, where $0<\beta^* \leq \beta$ and $\beta$ is the sampling budget. The computation time of actual distances between all sampled nodes is $O(\beta^*mn+\beta^*n^2)$. For $\beta^*=20\%$, only measuring the actual distances between sampled nodes will bring a $80\%$ reduction in computation time.\\

However, according to some approximation methods for SPLs discussed in section 3.3 and section 3.4, we can approximate the actual SPLs between sampled nodes instead of actually measuring them. And by doing that we can achieve further reduction in computation time. In the following section we will revise the approximation methods from \cite{ribeiro2012multiple} and \cite{potamias2009fast} and apply them to our random walk samples.\\

1) For networks with large $c.v.$, approximate actual SPLs by observed SPLs in the induced subgraph.\\

Based on theoretical and simulation results from \cite{ribeiro2012multiple}, in scale-free networks, random walks have strong ability to uncover the true shortest paths, so the actual SPLs between sampled nodes can be approximated by the their observed SPLs in the subgraph induced by the random walk sample. More specifically, for a pair of sampled nodes $(i, j)$, the actual SPL $l_{ij}$ between them in the population graph $G$ can be approximated by the observed SPL in the induced subgraph $G^*$.

More generally, it is the existence of hubs in scale-free networks that makes random walks able to find the shortest paths, as discussed in section 4.1. Therefore in this paper, we generalize the condition for random walks to uncover shortest paths to networks with relatively large variance in degree distribution, compared to the mean degree $<k>$. Let $c.v.=\frac{\sqrt{Var(k)}}{<k>}=\frac{\sqrt{<k^2>-<k>^2}}{<k>}$ denote the \textit{coefficient of variation} of the degree distribution as a measure of the relative variance. A large $c.v.$ is needed in order for the random walks to uncover the shortest paths, and we will discuss in section 5.1 about how large the $c.v.$ needs to be.\\

In an induced subgraph with $\beta^*n$ nodes, the computing time for single source shortest paths is reduced to $O(\beta^*m+\beta^*n)$ by BFS within the induces subgraph. Applying BFS to $\beta^*n$ sampled nodes in the induced subgraph, the time complexity for computing SPLs between all sampled nodes is $O(\beta^{*2}mn+\beta^{*2}n^2)$. Comparing to measuring the actual distance between sampled nodes, i.e., applying BFS to sampled nodes in the population graph, doing BFS only in the induced subgraph can save us $(1-\beta^*)\times 100 \%$ in computation time.\\

2) For networks with small $c.v.$, approximate actual SPLs using landmarks.\\

For networks with small $c.v.$ in degree distribution, since random walks can't find the shortest paths in the induced subgraph, we need to implement breadth-first search (BFS) on sampled nodes in the population graph to find the shortest paths. However, based on findings by \cite{potamias2009fast}, the BFS doesn't have to be applied to all sampled nodes. Instead, one can apply BFS  to only a fraction of the sampled nodes to find their shortest distances to all other nodes, and use that information to estimate the shortest distances between other sampled nodes. More specifically, one can first select a set of nodes as landmarks, denoted as $D$, pre-compute the SPLs from landmarks to all other nodes by BFS in the population graph, and estimate the SPL between any arbitrary pair of nodes $s$ and $t$ by $min_{j \in D}\{l_{sj}+l_{jt}\}$. The estimation will be very precise if many shortest paths contain the selected landmarks. From their experiments, using $100$ nodes with highest degrees from the population seems a fairly good strategy for choosing landmarks.\\

In this paper, we propose selecting landmarks from the sample. This is because we are only interested in the SPLs between nodes in the sample, and landmarks from the sample will be more likely to be on the shortest paths between nodes in the sample. Also it is costly to select landmarks from the population since we need to observe the degrees of all nodes. Let $\gamma$ denote the the ratio of number of landmarks to number of nodes in the induced subgraph $G^*$. From the sample we will choose the top $\gamma \beta^* n$ nodes in their actual degrees as landmarks. We will discuss the size of landmark set, i.e., the value of $\gamma$, in section 5.2.\\

In an induced subgraph with $\beta^*n$ nodes and $\gamma \beta^*n$ landmarks, the computing time for SPLs between a single landmark and all other nodes in the sample is still $O(m+n)$, since the BFS needs to be implemented in the population graph to compute the actual distances. Invoking the BFS $\gamma \beta^*n$ times, the computing time for SPLs between all landmarks and all other nodes in the sample is $O(\gamma \beta^* mn+\gamma \beta^*n^2)$. Comparing to measuring the actual distance between sampled nodes, i.e., applying BFS to all sampled nodes in the population graph, doing BFS only to the landmarks can save us $(1-\gamma)\times 100 \%$ in computation time. This is for the pre-computing stage. \\

For the estimation stage, for any arbitrary pair of nodes, it only takes $O(\gamma \beta ^*n)$ time to go through the distances from these two nodes to each landmark and choose the minimum sum as the estimated SPL. Note that with BFS applied to landmarks, the distances between $\gamma \beta ^*n$ landmarks and all other nodes in the sample have already been identified, therefore we just need to estimate the distances between $(1-\gamma)\beta ^*n$ nodes that are not used as landmarks. Applying $\gamma \beta ^*n$ numerical search to $\binom{(1-\gamma)\beta^*n}{2} \approx \frac{1}{2}(1-\gamma)^2\beta^{*2}n^2$ pairs of nodes in the sample, the computing time for estimating distances between sampled nodes that are not landmarks is about $O(\frac{1}{2}\gamma (1-\gamma)^2\beta^{*3}n^3)$ after we have the pre-computation data.

\subsection{Application of Estimating Methods}
In practice, sometimes we are only able to crawl part of the network, so we are restricted to observing the degrees of the sampled nodes. To apply the estimators in section 4.4 to estimating the SPLD for a network, we need to estimate $\psi_r$'s and $\pi_r$'s of the sampled nodes and $c.v$ of degree distribution by the degrees of nodes in the sample.\\

Following the mathematical expressions of $c.v.$, $\psi_r$, and $\pi_r$, we can estimate them by the estimated first moment $<k>$ and the second moment $<k^2>$ of the degree distribution. The estimation for $<k>$ and $<k^2>$ can be achieved by Hansen-Hurwitz ratio estimator. Suppose a single random walk $\{X_t\}$ initiates from its stationary distribution $\boldsymbol{p} = (\frac{k_1}{K}, \frac{k_2}{K}, ..., \frac{k_n}{K})^T$ on a connected graph $G$ with at least one triangle such that
\begin{equation}
\phi_i=\frac{k_i}{K}=\frac{k_i}{n<k>}.
\end{equation}
Then we can estimate the first moment $<k>$ by 
\begin{equation}
\hat{k}_1 = \frac{\hat{K}}{\hat{n}} = \frac{\frac{1}{|s|}\sum_{i \in s}\frac{k_i}{\phi_i}}{\frac{1}{|s|}\sum_{i \in s}\frac{1}{\phi_i}}  = \frac{\frac{1}{|s|}\sum_{i \in s}\frac{k_i}{\frac{k_i}{K}}}{\frac{1}{|s|}\sum_{i \in s}\frac{1}{\frac{k_i}{K}}}=\frac{|s|}{\sum_{i \in s}k_i^{-1}}.
\end{equation}
Similarly, we can estimate the second moment $<k^2>$ by 
\begin{equation}
\hat{k}_2= \frac{\frac{1}{|s|}\sum_{i \in s}\frac{k_i^2}{\phi_i}}{\frac{1}{|s|}\sum_{i \in s}\frac{1}{\phi_i}}  = \frac{\frac{1}{|s|}\sum_{i \in s}\frac{k_i^2}{\frac{k_i}{K}}}{\frac{1}{|s|}\sum_{i \in s}\frac{1}{\frac{k_i}{K}}}=\frac{\sum_{i \in s}k_i}{\sum_{i \in s}k_i^{-1}}.
\end{equation}

\subsubsection{Estimation of $c.v.$}
We can estimate $c.v.$ by 
\begin{equation}
\hat{c.v.} = \frac{\sqrt{\hat{k}_2-(\hat{k}_1)^2}}{\hat{k}_1}.
\end{equation}

\subsubsection{Estimation of $\psi_r$}
For Hansen-Hurwitz estimator, we can estimate $\alpha$ in $\psi_r=\alpha k_ik_j$ by
\begin{equation}
\hat{\alpha} = \frac{2}{(n\hat{k}_1)^2-n\hat{k}_2}, 
\end{equation}
and can therefore estimate $\psi_r$ by
\begin{equation}
\hat{\psi}_r = \frac{2}{(n\hat{k}_1)^2-n\hat{k}_2}k_ik_j.
\end{equation}
Note that for Hansen-Hurwitz ratio estimator (\ref{HH.ratio}), we can just plug in the observed degrees $k_i$ and $k_j$ of sampled nodes, and don't need to estimate any selection probabilities.\\

\subsubsection{Estimation of $\pi_r$}
For Horvitz-Thompson estimator, we can estimate $\tau_i$ identified in (\ref{tau}) by
\begin{equation}
\hat{\tau}_i=\frac{|s^*|}{n\hat{\bar{\theta}}}\hat{\theta}_i,
\end{equation}
where
\begin{equation}
\hat{\theta}_i=1-(1-\frac{k_i}{n\hat{k}_1})^t
\end{equation}
and
\begin{equation}
\hat{\bar{\theta}}=\frac{\frac{1}{|s|}\sum_{i \in s}\frac{\hat{\theta}_i}{\phi_i}}{\frac{1}{|s|}\sum_{i \in s}\frac{1}{\phi_i}}=\frac{\frac{1}{|s|}\sum_{i \in s}\frac{\hat{\theta}_i}{k_i/K}}{\frac{1}{|s|}\sum_{i \in s}\frac{1}{k_i/K}}=\frac{\sum_{i \in s}\frac{\hat{\theta}_i}{k_i}}{\sum_{i \in s}\frac{1}{k_i}}.
\end{equation}
Consequently, we can estimate $\pi_r$ by
\begin{equation}
\hat{\pi}_r=\hat{\tau}_i\hat{\tau}_j.
\end{equation}

\subsection{Evaluation Techniques}
To evaluate the performance of an estimator, we take $K$ random walk samples from the population graph $G$, compute the  estimate from each sample, and then apply the following four evaluating techniques to get an overall assessment for the estimator.

\subsubsection{Box plots}
\begin{figure}[H]
\begin{center}
\includegraphics[scale=0.5]{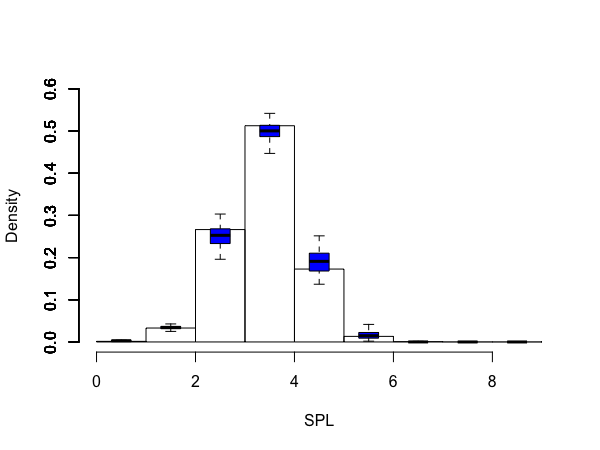}
 \caption{Box plots of estimated SPLDs on the histogram of population SPLD.}
 \label{boxplot}
\end{center}
\end{figure}

We first plot the histogram of the population SPLD. For each value of the population SPL, we place a box plot of sample estimates on the corresponding position of the histogram. Figure \ref{boxplot} is an example of box plots of Hansen-Hurwitz ratio estimates based on 100 samples taken from a scale-free network of size 1000. For each sample, a single random walk of 200 steps is used to produce the induced subgraph for the sample SPL to be observed.

\subsubsection{Mean Absolute Difference (MAD)}
For each value of population SPL $l$, the Mean Absolute Difference (MAD) for the estimated fraction $\hat{P}(l)$ is
\begin{equation}
mad(l) = E(|\hat{P}(l)-P(l)|).
\end{equation}
The empirical MAD for SPL $l$ from $K$ samples is
\begin{equation}
MAD(l) = \frac{1}{K}\sum_{k}|\hat{P}_k(l)-P(l)|,
\end{equation}
with estimated variance 
\begin{equation}
\hat{Var}(MAD(l)) = \frac{1}{K}\frac{\sum_{k}(|\hat{P}_k(l)-P(l)|-MAD(l))^2}{K-1}.
\end{equation}

Averaging all possible values of population SPL, the MAD for the estimated SPLD $\hat{P}$ is 
\begin{equation}
MAD = \frac{1}{L}\sum_{l}MAD(l),
\end{equation}
with estimated standard error
\begin{equation}\
\hat{se}(MAD) = \frac{1}{L}\sqrt{\sum_{l}\hat{Var}(MAD(l))}
\end{equation}

\subsubsection{Root Mean Square Error (RMSE)}
For each value of population SPL $l$, the Root Mean Square Error (RMSE) for the estimated fraction $\hat{P}(l)$ is
\begin{equation}
rmse(l) = \sqrt{E[(\hat{P}(l)-P(l))^2]}.
\end{equation}
The empirical RMSE for SPL $l$ from $K$ samples is
\begin{equation}
RMSE(l) = \sqrt{\frac{1}{K}\sum_{k}(\hat{P}_k(l)-P(l))^2},
\end{equation}
with estimated variance
\begin{equation}
\hat{Var}(RMSE(l)) = \frac{1}{K}\frac{\sum_{k}(\sqrt{(\hat{P}_k(l)-P(l))^2}-RMSE(l))^2}{K-1}.
\end{equation}

Averaging all possible values of population SPL, the RMSE for the estimated SPLD $\hat{P}$ is 
\begin{equation}
RMSE = \frac{1}{L}\sum_{l}RMSE(l),
\end{equation}
with estimated standard error
\begin{equation}
\hat{se}(RMSE) = \frac{1}{L}\sqrt{\sum_{l}\hat{Var}(RMSE(l))}.
\end{equation}
 
\subsubsection{Kullback-Leibler Divergence}
To measure the difference between two discrete distributions: estimated SPLD $\hat{P}_k$ from the $k^{th}$ sample, and population SPLD $P$, we can use the symmetrised Kullback-Leibler divergence:
\begin{equation}
KL(k)= \sum_{l}\hat{P}_k(l) log \frac{\hat{P}_k(l)}{P(l)} + \sum_{l}P(l) log \frac{P(l)}{\hat{P}_k(l)}.
\end{equation}
The average Kullback-Leibler divergence over all $K$ samples is
\begin{equation}
KL =\frac{1}{K}\sum_{k}KL(k),
\end{equation}
with estimated standard error
\begin{equation}
\hat{se}(KL) = \sqrt{\frac{1}{K}\frac{\sum_{k}(KL(k)-KL)^2}{K-1}}
\end{equation}

In practice, since the values of $KL$ are much almost ten times as large as the values of $MAD$ and $RMSE$, we will use $KL/10$ to keep the three numerical measures in the same scale.

\section{Simulation Study}
In this section, we present several simulation studies to assess the performance of the methods we proposed in Section 4. More specifically, by using the evaluation techniques discussed in section 4.6, we 1) test on different values of $c.v.$ of degree distribution to explore the conditions for random walks to uncover shortest paths; 2) test on various lengths and numbers of random walks and different estimators to find the best sampling design; 3) compare our estimates based on approximated SPLs to the unweighted sample SPLDs and estimates based on actual SPLs to evaluate the estimation performance.

\subsection{Conditions for Random Walks to Uncover Shortest Paths}
In Section 4.4.4, we generalized the condition for random walks to uncover shortest paths to having a large $c.v.$ of degree distribution. In this section, we will first verify the strong ability of random walks from scale-free networks in uncovering shortest paths. And based on that, we will explore the range of $c.v.$ which allows the random walks to perform well in uncovering shortest paths in general networks. To assess the performance, we will look at the proportion of shortest paths uncovered by the random walk sample. We will use networks with gamma degree distributions as an example of general networks.\\

In addition, as discussed by \cite{ribeiro2012multiple}, in networks with large degree variability, the fraction of edges with at least one its endpoints visited by the random walk is large. In this paper, we are more concerned about the fraction of edges in the induced subgraph, i.e., with edges with both endpoints visited by the random walk, because they are what we use to measure sample SPLs. If more edges are included in the induced subgraph, it is more likely to observe the true shortest paths from the sample. Let $E.f$ denote the fraction of edges with both of its endpoints visited by the random walk, that is, the fraction of edges in the induced subgraph. One should expect large values of $E.f$ for networks with large value of $c.v.$\\

For each network of size 1000, a single random walk of 200 steps is implemented to produce the induced subgraph. For each dyad in the subgraph, we take the difference between its sample SPL (SPL observed in the induced subgraph) and population SPL (SPL observed in the population graph, i.e., true SPL). Note that the sample SPL is always as large as or larger than the population SPL, as a node may take more steps in the subgraph to reach another node than it would in the population graph. Therefore the value of this difference has a range $\{0 ,1, 2, ...\}$. For each value of population SPL, we plot the distribution of difference between sample SPL and population SPL. The proportion of uncovered shortest paths by the random walk sample is equal to the proportion of zero difference between sample SPL and population SPL. Therefore, we expect a large proportion with zero difference to show that the random walk sample is performing well in uncovering the true SPL.\\

\begin{figure}[H]
\centering
\begin{tabular}{cc}
  \includegraphics[scale=0.4]{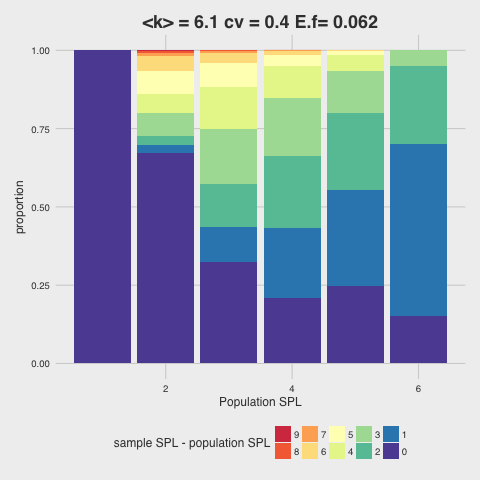} &   \includegraphics[scale=0.4]{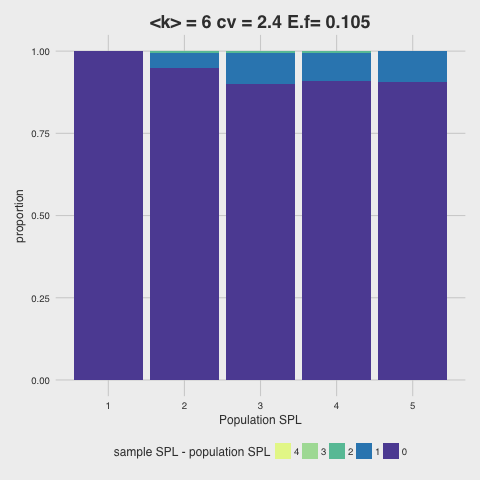} \\
(a) Erd\H{o}s-R\'enyi, $n=1000$, $\beta=0.2$ & (b) scale-free, $n=1000$, $\beta=0.2$\\[6pt]
\end{tabular}
\caption{Erd\H{o}s-R\'enyi network v.s. scale-free network: distribution of difference between sample SPL and population SPL.}
\label{ER_scale_free}
\end{figure}

1) Scale-free networks v.s. Erd\H{o}s-R\'enyi networks\\
We first compare a Erd\H{o}s-R\'enyi network and a scale-free network, both of which have average degree around 6. In Figure \ref{ER_scale_free}, we observe a large proportion of zero difference for each value of SPL in the scale-free network, which indicates that random walks have strong ability in uncovering the true shortest paths. However, in the Erd\H{o}s-R\'enyi network, we don't see a large proportion of zero difference, for any value of SPL greater than 1. Therefore the ability of random walks to uncover the true shortest paths in the Erd\H{o}s-R\'enyi is very weak. This is to be expected, since the $c.v.$ of degree distribution of the scale-free network is much larger than that of the Erd\H{o}s-R\'enyi network. What's more, we notice that $E.f$ in the scale-free network is larger than that in the Erd\H{o}s-R\'enyi network, which also explains why random walks are doing a better job in uncovering shortest paths in the scale-free network.\\

2) General Networks\\
A more general condition for random walks to uncover shortest paths is that the degree distribution has a large coefficient of variation ($c.v.$). To explore how large the $c.v.$ needs to be in order for the random walk to perform well in uncovering the shortest paths, we compare 4 networks with gamma degree distributions.\\

As one would expect, as the $c.v.$ increases from $0.8$ in network $(c)$ to $2.4$ in network $(f)$, $E.f$ increases, which means more edges are observed in the induced subgraph, and therefore the proportion of zero difference between sample SPL and population SPL increases. When $c.v.$ reaches $1.8$ in network $(e)$, the distribution of difference between sample SPL and population SPL looks very close to that for the scale-free network in Figure \ref{ER_scale_free}. When $c.v.$ increases from $1.8$ in network $(e)$ to $2.4$ in network $(f)$, there is still an increase in the proportion of zero difference between sample SPL and population SPL, but not very substantial. One should also notice that $c.v.$ for the scale-free network in Figure \ref{ER_scale_free} is $2.4$. Combining the empirical results from some real networks in section 6, we get some insight about the value of $c.v.$ we need for the random walk to perform well in uncovering shortest paths:\\

1) If the $c.v.$ is much smaller than 2, the random walk is not able to uncover the shortest paths;\\

2) If the $c.v.$ is around 2, the random walk has the ability to uncover the shortest paths, but the performance may vary from case to case;\\

3) If the $c.v$ is much larger than 2, the random walk has strong ability to uncover most of the shortest paths between the sampled nodes.\\

As network $(f)$ has the same value of $c.v.$ as the scale-free network $(b)$, we will use degree sequence generated from $Gamma(0.125,40)+1$ to generate networks as an example for networks with large $c.v.$ in the rest of this simulation section. And we will use degree sequence generated from $Gamma(1,5)+1$ (setting for network $(c)$) to generate networks as an example for networks with small $c.v.$. In order to evaluate the estimation performance, for a given network, a specific sampling design and a specific estimator, a total of $K=100$ random walk samples will be drawn from the network. An estimate will be computed from each of the samples. Then the $100$ estimates will be used to construct the the box plots and calculate the three numerical measures discussed in section 4.6.\\

\begin{figure}[H]
\centering
\begin{tabular}{c c}
      \includegraphics[scale=0.4]{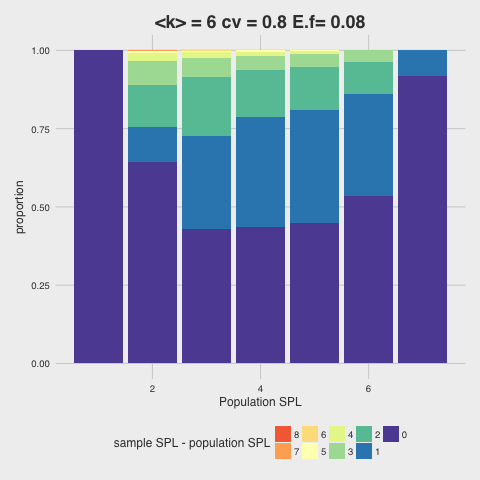} &
      \includegraphics[scale=0.4]{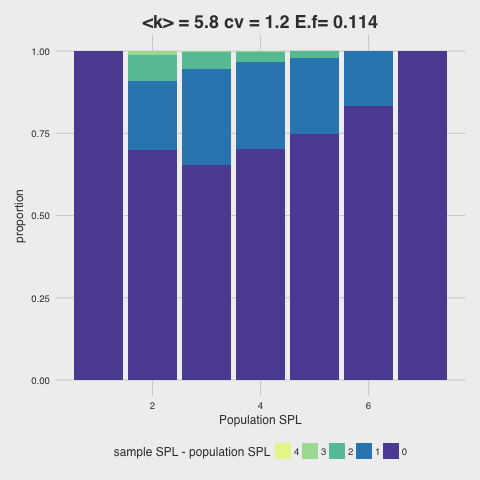} \\
      (c) Gamma(1,5)+1, $n=1000$, $\beta=0.2$ & (d) Gamma(0.5,10)+1, $n=1000$, $\beta=0.2$\\[6pt]
      \includegraphics[scale=0.4]{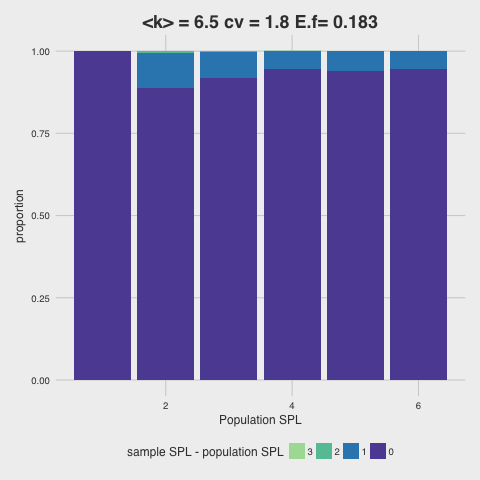} &
      \includegraphics[scale=0.4]{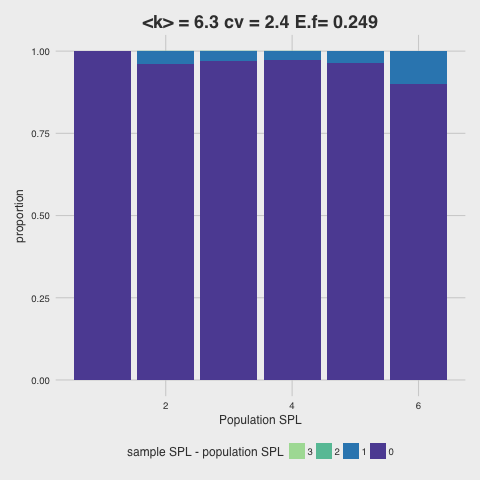} \\
      (e) Gamma(0.25,20)+1, $n=1000$, $\beta=0.2$ & (f) Gamma(0.125,40)+1, $n=1000$, $\beta=0.2$
\end{tabular}
\caption{Networks with Gamma degree distribution: distribution of difference between sample SPL and population SPL.}
\label{gamma}
\end{figure}

\subsection{Sampling designs for Random Walks}
In this section, we will explore random walk sampling designs for estimating the population SPLD. We will also compare the performance of different estimators. Basically, we will answer the following four questions:\\

1) For networks with large $c.v.$, how many steps do we need in a single random walk in order to get a good estimation? \\

2) For networks with small $c.v.$, how many nodes do we need to use as landmarks and how many steps do we need in a single random walk in order to get a good estimation?\\

3) Will multiple random walks outperform a single random walk, given fixed sampling budget?\\

4) For a fixed sampling design, how will the performance differ by using different estimators?

\subsubsection{Length of Random Walks for Networks with Large $c.v.$}

For networks with large $c.v.$ in degree distribution, we use the observed SPLs in the induced subgraph to approximate the actual SPLs between sampled nodes. In order to see the effect of length of a single random walk on the estimation performance, we implement single random walks with sampling budget $\beta = 0.05(0.05)0.5$, where $x = a(r)b$ means $x$ increasing from $a$ to $b$, with $r$ increment at each time. This process is applied to networks with $c.v.=2.4$ and size $n=1,000$, $n=5,000$, and $n=10,000$. The estimator we use here is the generalized Hansen-Hurwitz estimator, denoted as HH.ra.\\

\begin{figure}[H]
\centering
\begin{tabular}{c c}
\includegraphics[scale=0.4]{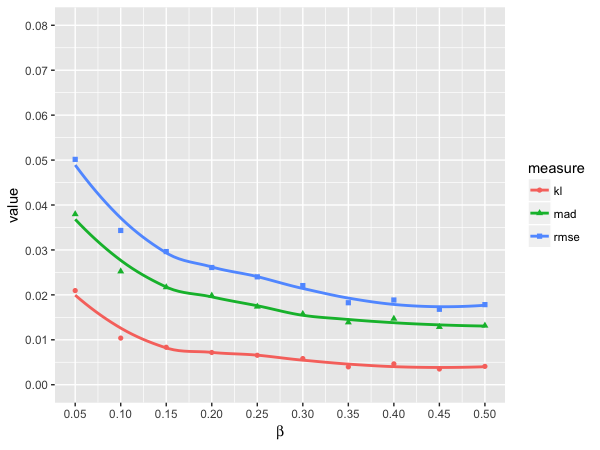}  
& \includegraphics[scale=0.4]{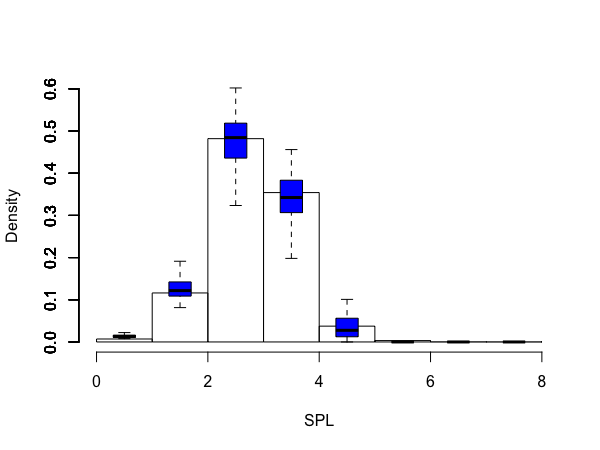}\\
(a) $n=1000$, $c.v.=2.4$ & (d) $n=1000$, $c.v.=2.4$, $\beta=0.2$\\
\includegraphics[scale=0.4]{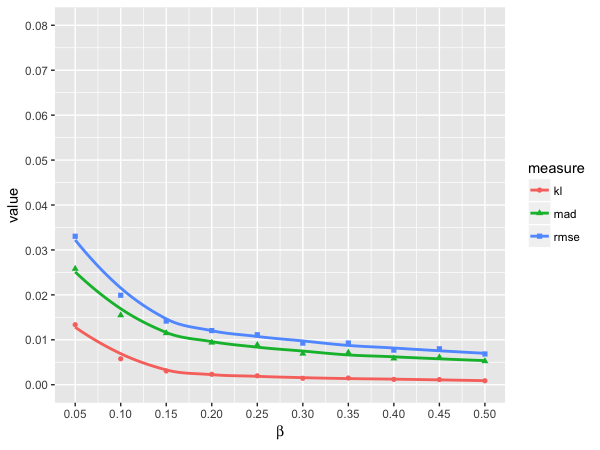}  
& \includegraphics[scale=0.4]{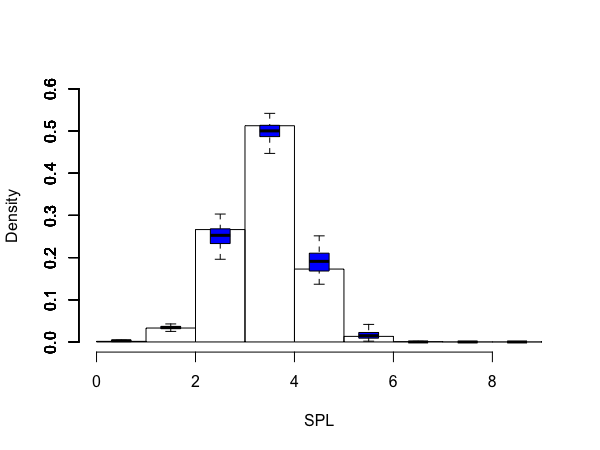}\\
(b) $n=5000$, $c.v.=2.4$ & (e) $n=5000$, $c.v.=2.4$, $\beta=0.2$\\
\includegraphics[scale=0.4]{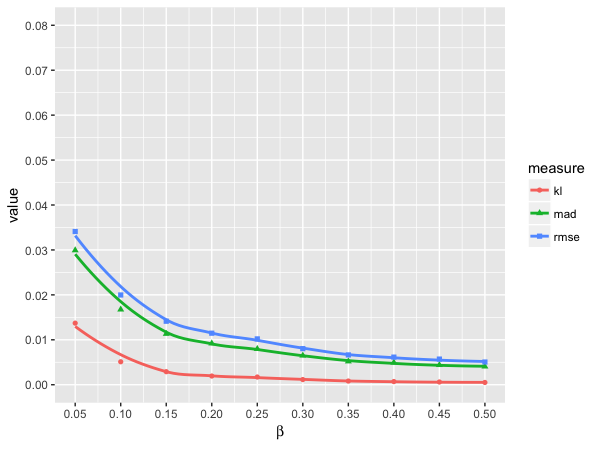}  
& \includegraphics[scale=0.4]{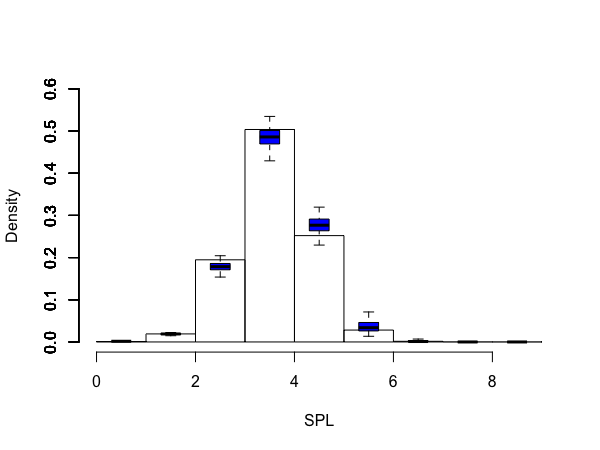}\\
(c) $n=10000$, $c.v.=2.4$ & (f) $n=10000$, $c.v.=2.4$, $\beta=0.2$
\end{tabular}
\caption{Performance of generalized Hansen-Hurwitz ratio estimator versus length ($\beta$) of random walks in networks with large $c.v.$, measured by $MAD$, $RMSE$, and $KL$ (low values are better).}
\label{cv>2_length}
\end{figure}

In Figure \ref{cv>2_length}, the values of the three numerical measures of accuracy keep decreasing, as we increase the sampling budget from $0.05$ to $0.5$. That means, the estimation performance is improving as the single random walk gets longer, which is to be expected. However, the improvement is dramatic as the sampling rate reaches $0.2$, and becomes moderate beyond that. Therefore, it is appropriate to set the minimum sampling budget $\beta$ to be around $0.2$ for the estimation to perform well. Let's now assume $\beta^*=\beta=0.2$, then the computing time of approximating SPLs between all sampled nodes is $O(0.04mn+0.04n^2)$. Comparing it to the computing time of actual distances between all sampled nodes $O(0.2mn +0.2n^2)$, approximating the SPLs leads to about $80\%$ reduction in computation.\\

\begin{figure}[H]
\centering
\begin{tabular}{c c}
\includegraphics[scale=0.4]{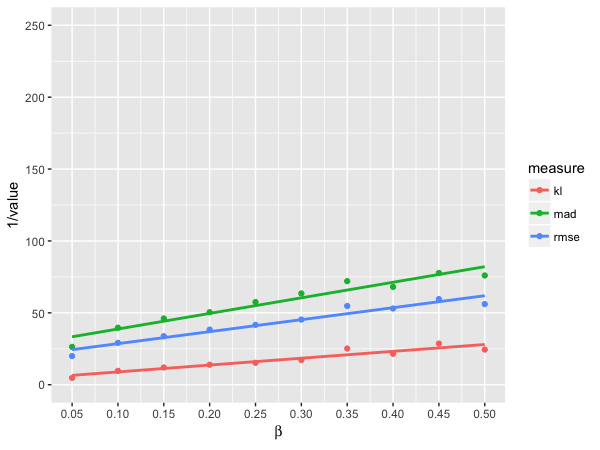}  
& \includegraphics[scale=0.4]{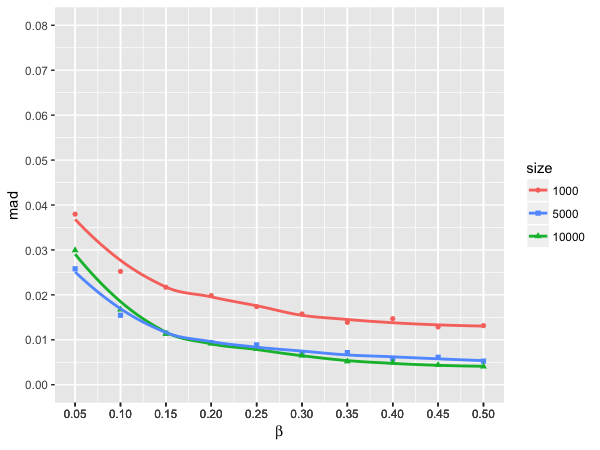}\\
(a) $n=1000$, $c.v.=2.4$ & (d) $MAD$, $c.v.=2.4$\\
\includegraphics[scale=0.4]{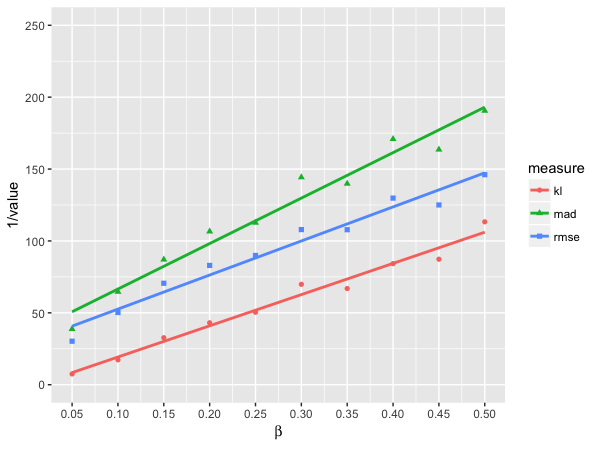}  
& \includegraphics[scale=0.4]{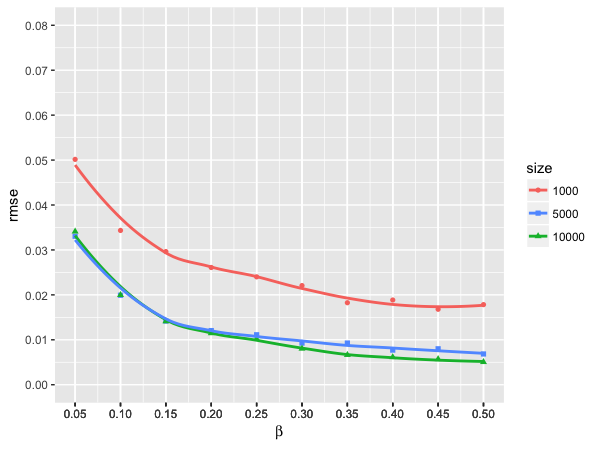}\\
(b) $n=5000$, $c.v.=2.4$ & (e) $RMSE$, $c.v.=2.4$\\
\includegraphics[scale=0.4]{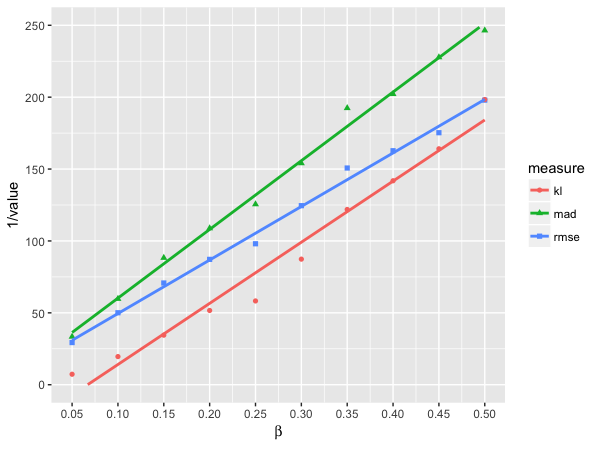}  
& \includegraphics[scale=0.4]{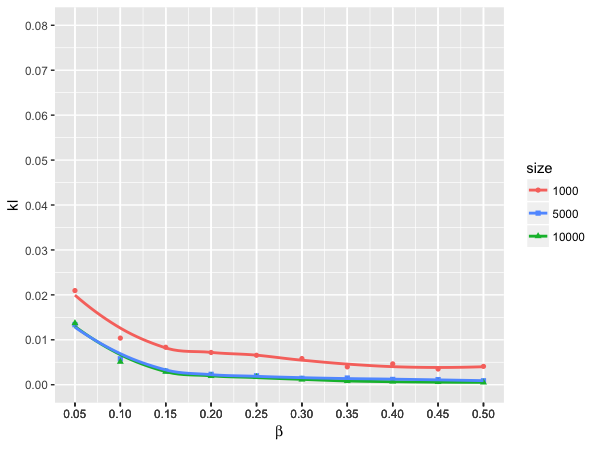}\\
(c) $n=10000$, $c.v.=2.4$ & (f) $KL$, $c.v.=2.4$
\end{tabular}
\caption{Performance of generalized Hansen-Hurwitz ratio estimator versus length ($\beta$) of random walks in networks with large $c.v.$ Left: comparison of performance measures under each network size, measured by the inverse of $MAD$, $RMSE$, and $KL$ (high values are better).  Right: comparison of network sizes under each performance measure, measured by $MAD$, $RMSE$, and $KL$ (low values are better).}
\label{cv>2_length_inverse}
\end{figure}

Another thing we can notice from Figure \ref{cv>2_length} is that the estimation performance is better in larger networks. More specifically, as we increase the network size, the estimates stay unbiased and their variance gets smaller. One possible reason for this phenomenon is the small world effect. For a fixed sampling budget, the sample size increases linearly with the network size, while the shortest path lengths only increases in the $\log$ scale. Therefore even with the same sampling budget, a random walk in a large network is relatively ``longer" than that in a small network, and thus has a stronger ability in uncovering the shortest paths. However, we can also observe from plots $(d)$, $(e)$ and $(f)$ in Figure \ref{cv>2_length_inverse} that the estimation performance for network of size $n=5000$ and that for network of size $n=10000$ are very similar. Therefore one can expect the relationship between estimation performance and sampling budget to be similar as plot $(c)$ in Figure \ref{cv>2_length} if a network with large $c.v.$ is of size $n=5000$ or larger.\\

What's more, as we can observe in plots $(a)$, $(b)$ and $(c)$ in Figure \ref{cv>2_length_inverse}, the inverse of the three estimation measures seem to have an approximately linear relationship with the sampling budget. If the coefficients of this linear relationship can be found for large networks, we can estimate the estimation accuracy in advance based on sampling budget.

\subsubsection{Size of Landmarks and Length of Random Walks for Networks with Small $c.v.$}

For networks with small $c.v.$ in degree distribution, due to the lack of powerful hubs, random walks lack strong ability to uncover shortest paths. As discussed in section 4.4.4, an alternative way is to use landmarks to estimate the SPLs between sampled nodes. We proposed using nodes in the sample with high degrees as landmarks, and a remaining question is the size of landmark set.\\

In order to see the effect of landmark size and single random walk length on the estimation performance, we will:\\

1) Fix the sampling budget at $\beta=0.2$ and let $\gamma = 0.05(0.05)0.5$ to find the minimum fraction $\gamma_0$ for good estimation;\\

2) Fix the fraction of landmarks at $\gamma=\gamma_0$, implement single random walks with sampling budget $\beta = 0.05(0.05)0.5$, and check if a random walk with $\beta<0.2$ is also acceptable.\\

\begin{figure}[H]
\centering
\begin{tabular}{c c}
\includegraphics[scale=0.4]{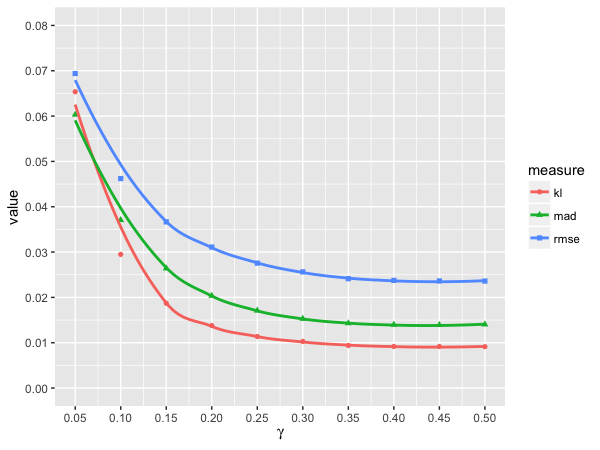}  
& \includegraphics[scale=0.4]{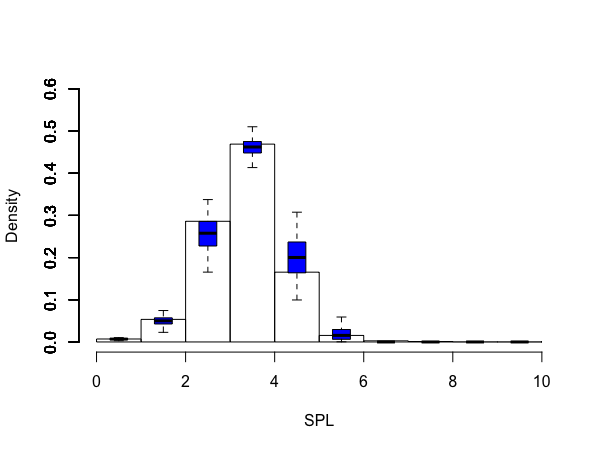}\\
(a) $n=1000$, $c.v.=0.8$, $\beta=0.2$ & (d) $n=1000$, $c.v.=0.8$, $\beta=0.2$, $\gamma=0.3$\\
\includegraphics[scale=0.4]{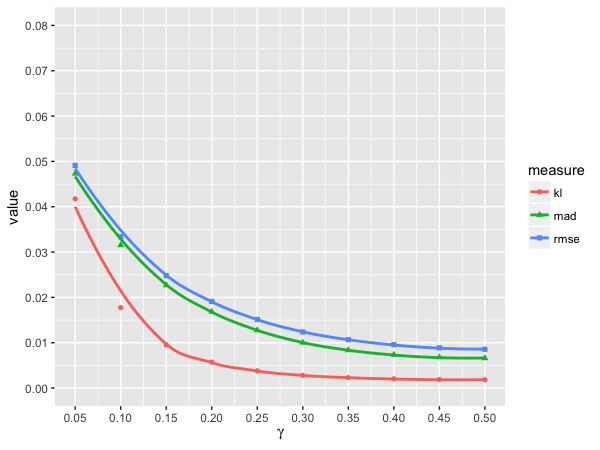}  
& \includegraphics[scale=0.4]{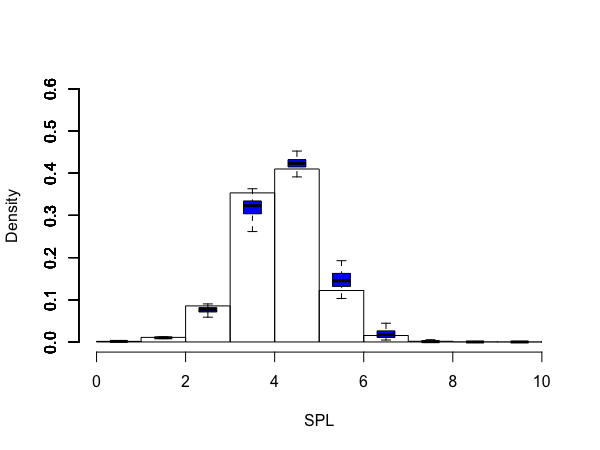}\\
(b) $n=5000$, $c.v.=0.8$, $\beta=0.2$ & (e) $n=5000$, $c.v.=0.8$, $\beta=0.2$, $\gamma=0.3$\\
\includegraphics[scale=0.4]{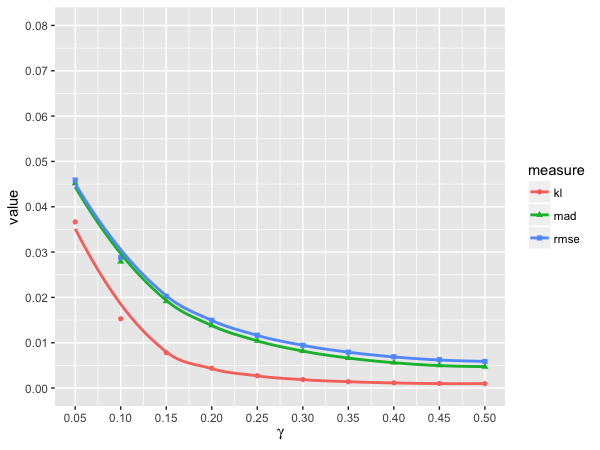}  
& \includegraphics[scale=0.4]{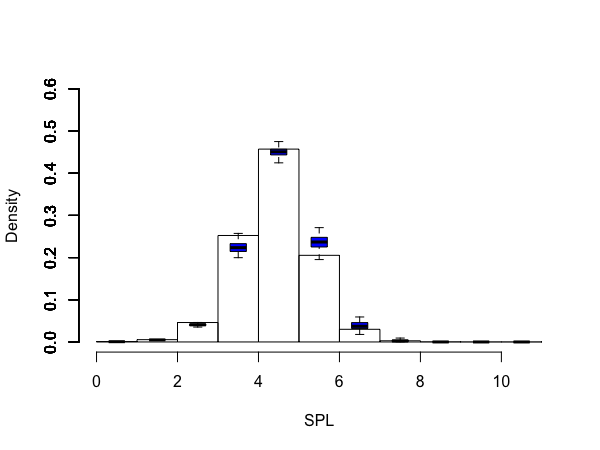}\\
(c) $n=10000$, $c.v.=0.8$, $\beta=0.2$ & (f) $n=10000$, $c.v.=0.8$, $\beta=0.2$, $\gamma=0.3$
\end{tabular}
\caption{Performance of generalized Hansen-Hurwitz ratio estimator versus size ($\gamma$) of landmarks in networks with small $c.v.$, measured by $MAD$, $RMSE$ and $KL$ (low values are better).}
\label{cv<2_gamma_length}
\end{figure}

\begin{figure}[H]
\centering
\begin{tabular}{c c}
\includegraphics[scale=0.4]{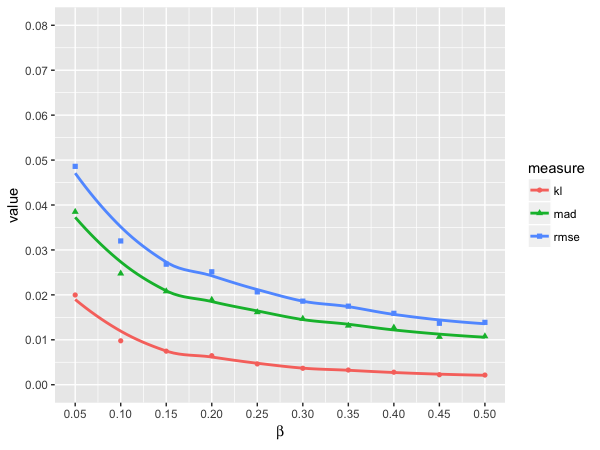}  
& \includegraphics[scale=0.4]{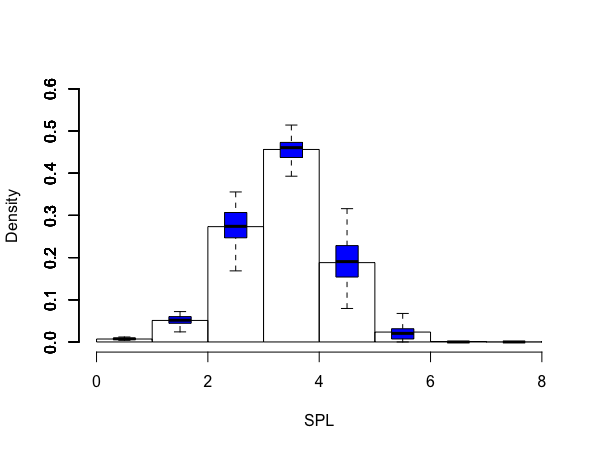}\\
(a) $n=1000$, $c.v.=0.8$, $\gamma=0.3$ & (d) $n=1000$, $c.v.=0.8$, $\gamma=0.3$, $\beta=0.2$\\
\includegraphics[scale=0.4]{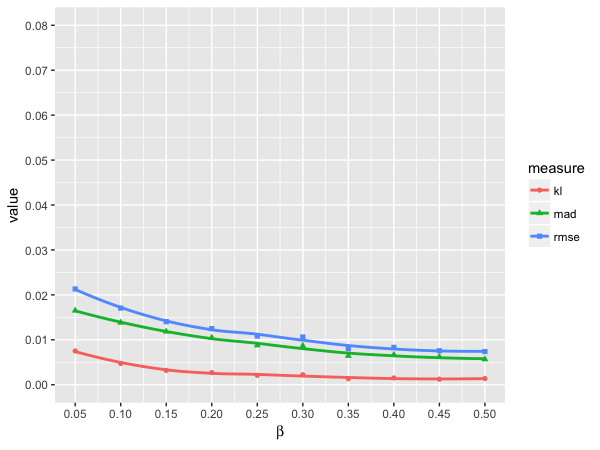}  
& \includegraphics[scale=0.4]{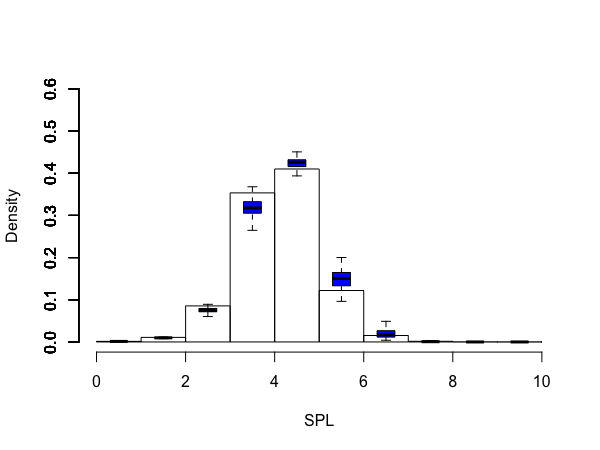}\\
(b) $n=5000$, $c.v.=0.8$,  $\gamma=0.3$ & (e) $n=5000$, $c.v.=0.8$, $\gamma=0.3$, $\beta=0.2$\\
\includegraphics[scale=0.4]{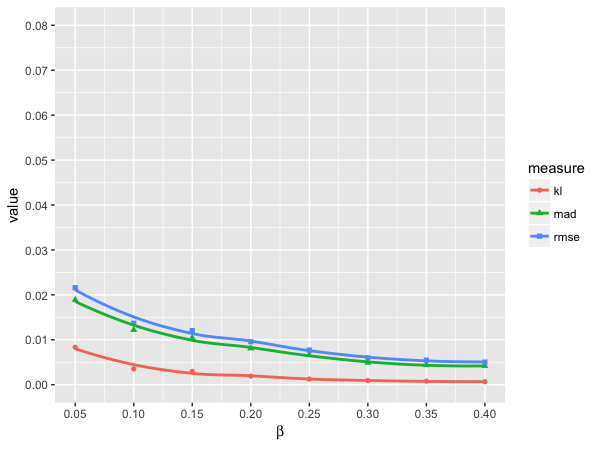}  
& \includegraphics[scale=0.4]{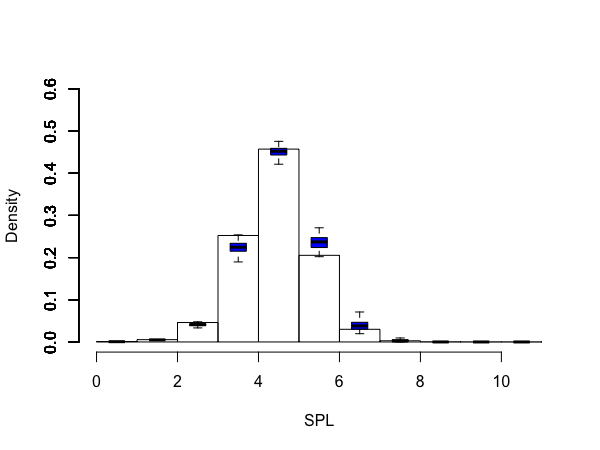}\\
(c) $n=10000$, $c.v.=0.8$,  $\gamma=0.3$ & (f) $n=10000$, $c.v.=0.8$, $\gamma=0.3$, $\beta=0.2$
\end{tabular}
\caption{Performance of generalized Hansen-Hurwitz ratio estimator versus length ($\beta$) of random walks in networks with small $c.v.$, measured by $MAD$, $RMSE$, and $KL$ (low values are better).}
\label{cv<2_beta_length}
\end{figure}

The above process is applied to networks with $c.v.=0.8$ and size $n=1,000$, $n=5,000$, and $n=10,000$, as shown in Figure \ref{cv<2_gamma_length} and \ref{cv<2_beta_length}. The estimator we use here is the generalized Hansen-Hurwitz estimator, denoted as HH.ra.\\

In Figure \ref{cv<2_gamma_length}, the values of the three numerical measures are decreasing as $\gamma$ increases from $0.05$ to $0.2$, and stay almost stable after $0.3$. Thus we can use $\gamma_0=0.3$ as the minimum fraction of landmarks. In Figure \ref{cv<2_beta_length}, for large networks when $n=5,000$ or $n=10,000$, the estimation performance is very good if we use a sampling budget as large as $\beta=0.2$. We can also use s smaller sampling budget such as $0.15$ or even $0.1$ for large networks since the estimation error will not increase too much. If we assume $\beta^*=\beta=0.2$ and use $\gamma=0.3$, the pre-computing time of approximating SPLs between all sampled nodes is $O(0.06mn+0.06n^2)$. Comparing it to the computing time of actual distances between all sampled nodes $O(0.2mn +0.2n^2)$, approximating the SPLs leads to about $70\%$ reduction in computation.\\

Similar to networks with large $c.v.$, for networks with small $c.v.$ we also notice that the estimation performance is better in larger networks. A possible reason is that as we increase the network size and fix sampling budget and landmark fraction, the number of landmarks is getting larger. And with more landmarks it is more likely to get a precise estimation of the SPLs between sampled nodes.\\

\begin{figure}[H]
\centering
\begin{tabular}{cc}
\includegraphics[scale=0.4]{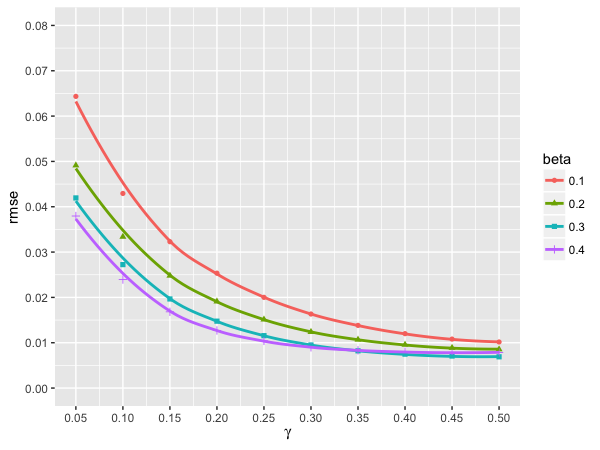}&
\includegraphics[scale=0.4]{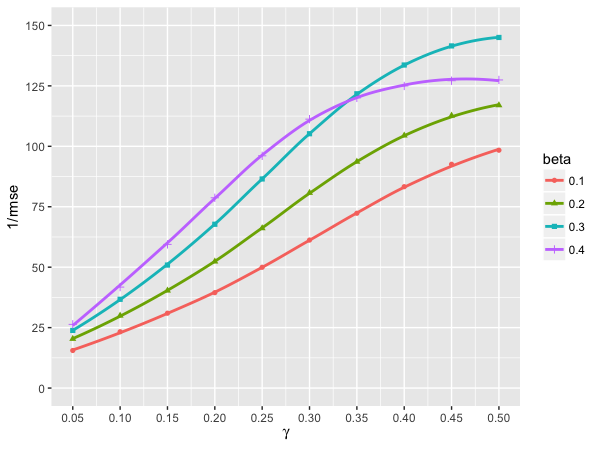}\\
(a) $RMSE$ & (b) $1/RMSE$
\end{tabular}
\caption{Performance of generalized Hansen-Hurwitz ratio estimator versus size ($\gamma$) of landmarks for different lengths ($\beta$) of random walks in a network with $n=5000$ and $c.v.=0.8$ (small).}
\label{gamma_beta}
\end{figure}

On the other hand, Figure \ref{gamma_beta} shows the change of RMSE as we increase the landmark size $\gamma$ for different values of random walk length $\beta$. As expected, the lines for larger $beta$ are below the lines for smaller $\beta$. The means if the random walk is longer, less landmarks are needed. To save computation time of breadth-first search, we want the value of $\beta \gamma$ to be as small as possible. The questions remains whether to use large $\beta$ and small $\gamma$ or to use small $\beta$ and large $\gamma$. Ideally the latter is better because by doing that we can also save the sampling cost. Suppose we want the RMSE to be as small as $0.01$, there are four available combinations of $\beta$ and $\gamma$ listed in Table \ref{comb} to achieve this accuracy. Among them $\beta=0.1$ and $\gamma=0.5$ is the best because it achieves both the smallest sampling budget and the shortest computation time for BFS.

\begin{figure}[H]
\centering
\begin{tabular}{c c}
\includegraphics[scale=0.4]{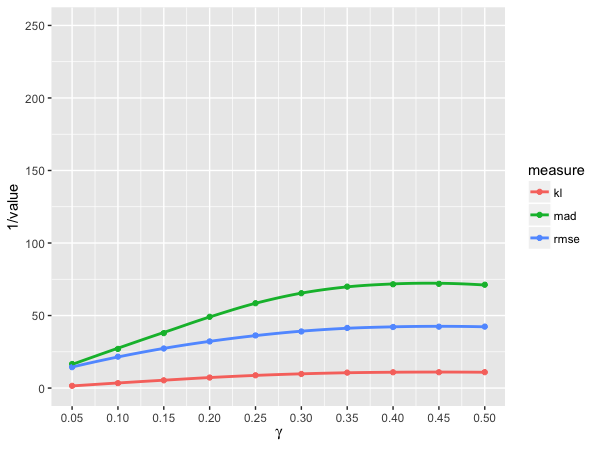}  
& \includegraphics[scale=0.4]{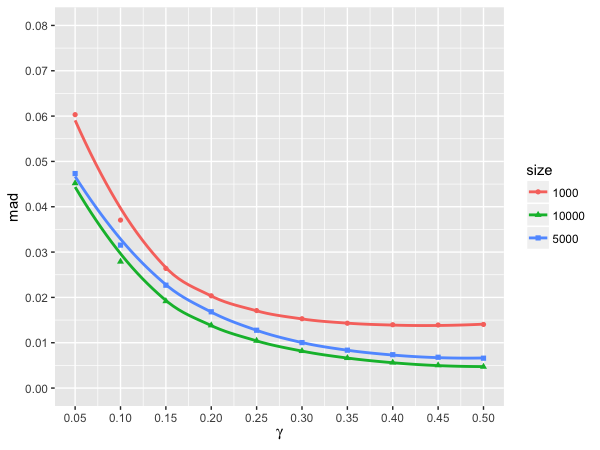}\\
(a) $n=1000$, $c.v.=0.8$, $\beta=0.2$ & (d) $MAD$, $c.v.=0.8$, $\beta=0.2$ \\
\includegraphics[scale=0.4]{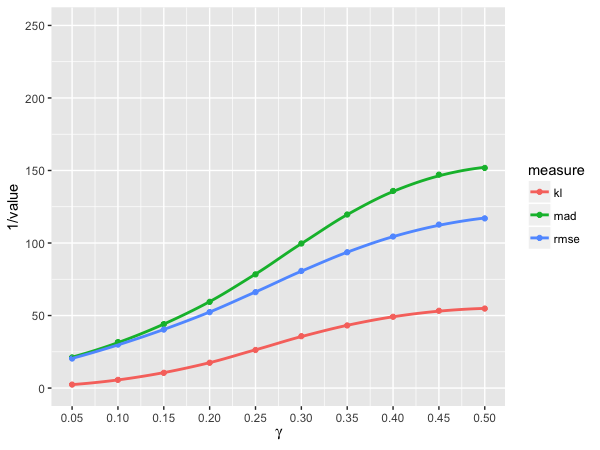}
& \includegraphics[scale=0.4]{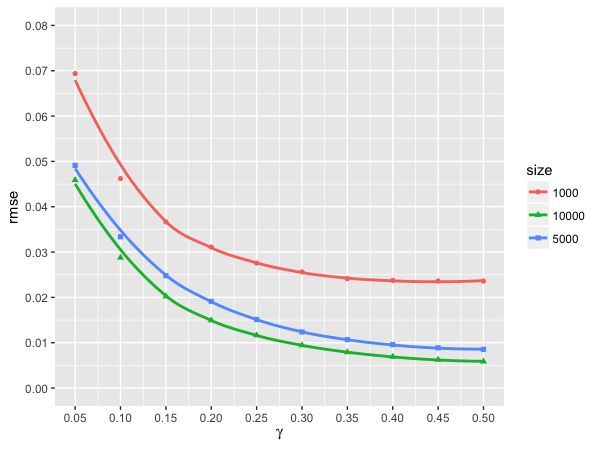}\\
(b) $n=5000$, $c.v.=0.8$, $\beta=0.2$ & (e) $RMSE$, $c.v.=0.8$, $\beta=0.2$ \\
\includegraphics[scale=0.4]{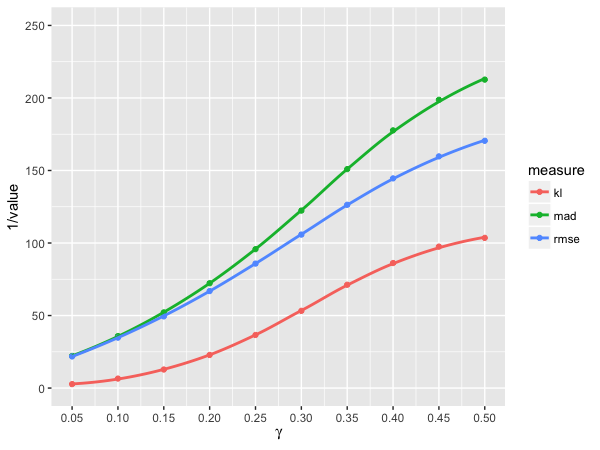}  
& \includegraphics[scale=0.4]{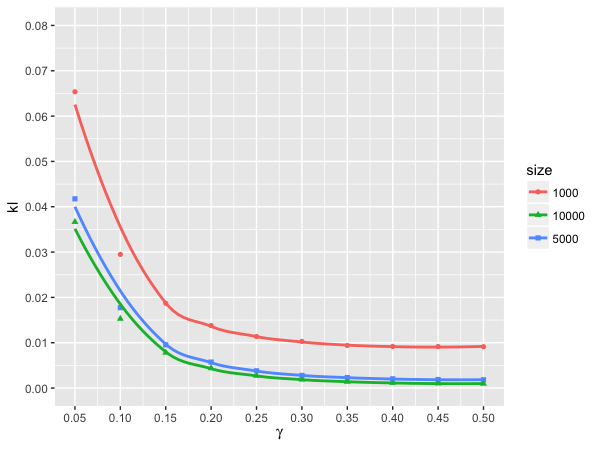}\\
(c) $n=10000$, $c.v.=0.8$, $\beta=0.2$ & (f) $KL$, $c.v.=0.8$, $\beta=0.2$ 
\end{tabular}
\caption{Performance of generalized Hansen-Hurwitz ratio estimator versus size ($\gamma$) of landmarks in networks with small $c.v.$ Left: comparison of performance measures under each network size, measured by the inverse of $MAD$, $RMSE$, and $KL$ (high values are better).  Right: comparison of network sizes under each performance measure, measured by $MAD$, $RMSE$, and $KL$ (low values are better).}
\label{cv<2_gamma_length_inverse}
\end{figure}

\begin{table}[H]
\centering
\begin{tabular}{c|c|c}
$\beta$ & $\gamma$ & $\beta\gamma$\\
\hline
0.4 & 0.25 & 0.1\\
0.3 & 0.3 & 0.09\\
0.2 & 0.375 & 0.075\\
0.1 & 0.5 & 0.05
\end{tabular}
\caption{Comparison of combinations of random walk length ($\beta$) and landmark size ($\gamma$) to achieve $RMSE \approx 0.01$ in a network with $n=5000$ and $c.v.=0.8$ (small).}
\label{comb}
\end{table}

\subsubsection{Number of Random Walks}
To compare the estimation performance with a single random walk and multiple random walks, we fix the total sampling budget and take $H$ independent random walk samples with $H$ ranging from $1$ to $6$. For networks with large $c.v.$, we fix the total sampling budget at $\beta_0=0.2$. For networks with small $c.v.$, we fix the total sampling budget at $\beta_0=0.2$ and use $\gamma_0=0.3$ as the landmark fraction.\\

\begin{figure}[H]
\centering
\begin{tabular}{c c}
\includegraphics[scale=0.4]{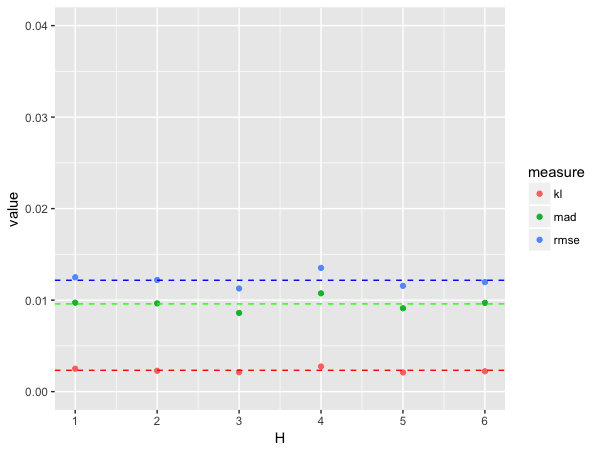}  
& \includegraphics[scale=0.4]{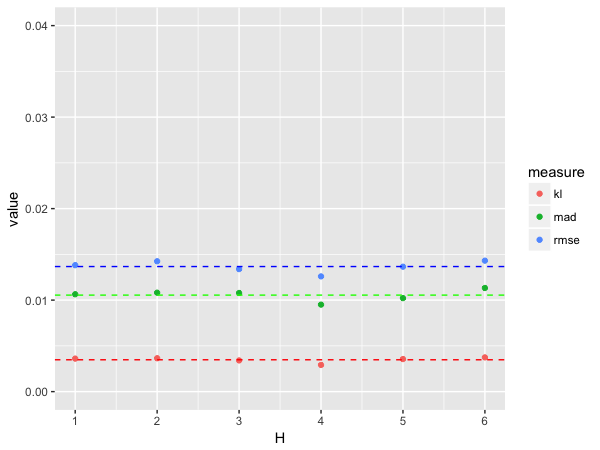}\\
(a) $n=5000$, $c.v.=2.4$, $\beta=0.2$  & (b) $n=5000$, $c.v.=0.8$, $\beta=0.2$, $\gamma=0.3$  \\
\end{tabular}
\caption{Performance of generalized Hansen-Hurwitz ratio estimator versus number ($H$) of random walks, measured by $MAD$, $RMSE$, and $KL$ (low values are better).}
\label{number}
\end{figure}

As we can observe in Figure \ref{number}, for both networks, the three numerical measures are stable as we increase the number of random walks from $1$ to $6$. Therefore, when keeping the total sampling budget fixed, using multiple random walks will not improve the estimation performance. In the case of networks with large $c.v$, the reason for this phenomenon is explained by \cite{ribeiro2012multiple}. As they have shown in their work, if the network has a large variance in degree distribution, two random walks intersect with high probability, and thus the subgraph induced by multiple random walks will be very similar to that induced by a single random walk. In the case of networks with small $c.v.$, where we use landmarks to estimate the SPLs between sampled nodes, although the landmarks found by a single random walk and those by multiple random walks are not necessarily the same, our simulation showed that they have similar and high betweenness centralities. We can therefore infer that they will play similar roles in estimating the distances between other nodes. 
\\

\subsubsection{Comparison of Estimators}

\begin{figure}[H]
\centering
\begin{tabular}{c c}
\includegraphics[scale=0.4]{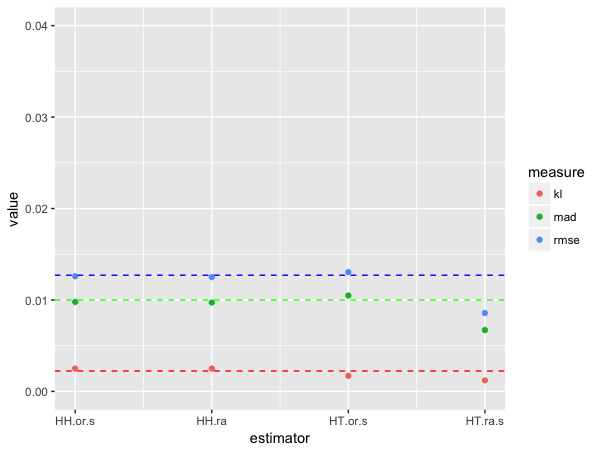}  
& \includegraphics[scale=0.4]{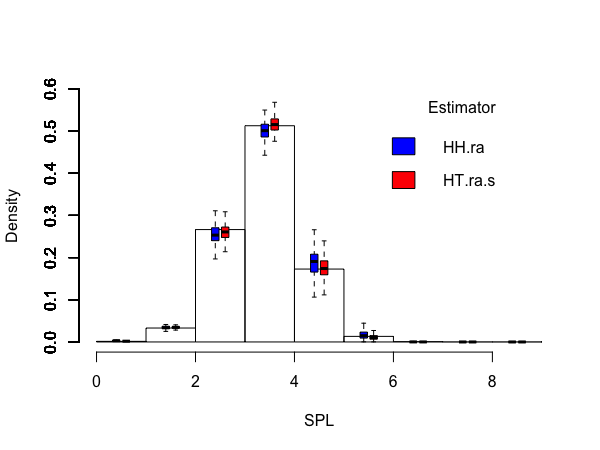}\\
(a) $n=5000$, $c.v.=2.4$, $\beta=0.2$  & (b) $n=5000$, $c.v.= 2.4$, $\beta=0.2$\\
\includegraphics[scale=0.4]{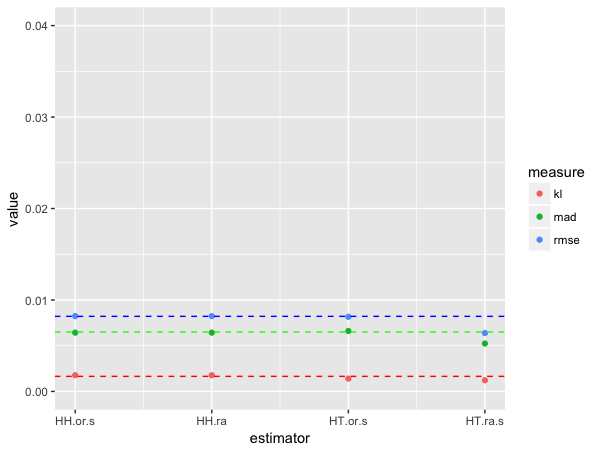}  
& \includegraphics[scale=0.4]{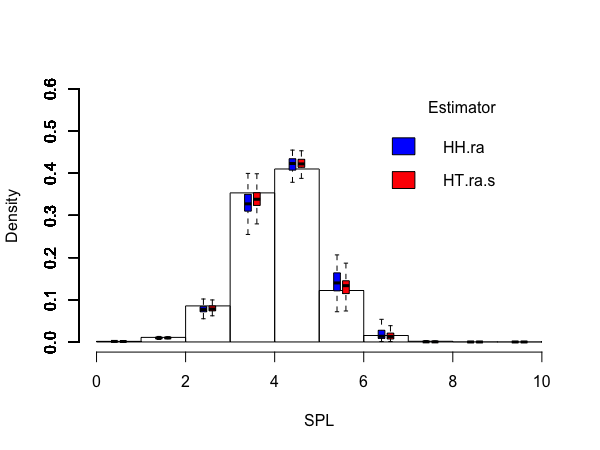}\\
(c) $n=5000$, $c.v.=0.8$, $\beta=0.2$, $\gamma=0.3$  & (d) $n=5000$, $c.v.= 0.8$, $\beta=0.2$, $\gamma=0.3$
\end{tabular}
\caption{Estimation performance versus estimators, measured by $MAD$, $RMSE$, and $KL$ (low values are better).}
\label{HH_HT}
\end{figure}

In this section, we compare the performances of the four estimators proposed in section 4.4. For generalized Hansen-Hurwitz estimator, Horvitz-Thompson estimator, and Horvitz-Thompson ratio estimator, $\psi_r$'s and $\pi_r$'s are estimated by the expressions discussed in section 4.5, therefore the estimates are denoted by HH.or.s, HT.or.s, and HT.ra.s, respectively. For generalized Hansen-Hurwitz ratio estimator, we just need to use the actual degrees of sampled nodes to compute the estimates, thus the estimates are denoted as HH.ra. The comparison based on numerical evaluations measures and comparison based on box plots are shown in Figure \ref{HH_HT}.\\

From the numerical comparison, one can observe that the Horvitz-Thompson ratio estimator is doing a slightly better job than the other three estimators. As we can observe from the comparison of box plots, the Horvitz-Thompson ratio estimator exhibits smaller variance than the Hansen-Hurwitz ratio estimator. There are two reasons for this phenomenon. According to the Rao-Blackwell theorem (\cite{casella2002statistical}, p.342), if $\hat{\theta}$ is an unbiased estimator of $\theta$ and $\theta^* = E(\hat{\theta}|T)$ where $T$ is the sufficient statistic for $\theta$, then $\theta^*$ is also an unbiased estimator of $\theta$ and $Var(\theta^*) \leq Var(\hat{\theta})$, and the inequality is strict unless $\theta$ is a function of $T$. That is, for any unbiased estimator that is not a function of the sufficient statistic, one may always obtain an unbiased estimator, depending on the sufficient statistic, that is better in terms of smaller variance. For the finite population sampling situation, the \textit{minimal sufficient statistic $T$} is the unordered set of distinct, labeled observations (\cite{basu1969role}). Therefore, the Hansen-Hurwitz estimator $\hat{t}^{HH}$ is not a function of the minimal sufficient statistic while the Horvitz-Thompson estimator $\hat{t}^{HT}$ is. Note that both $\hat{t}^{HH}$ and $\hat{t}^{HT}$ are unbiased estimators for $t$. Based on the Rao-Blackwell theorem, we can always find another unbiased estimator $W=E(\hat{t}^{HH}|T)$ such that $W$ has a smaller variance than $\hat{t}^{HH}$, while we cannot find such an estimator for $\hat{t}^{HT}$ as $\hat{t}^{HT} = E(\hat{t}^{HT}|T)$. Therefore $\hat{t}^{HT}$ is expected to have a smaller variance than $\hat{t}^{HH}$. Second, since the ratio form ensures that the estimated fractions for all values of SPL sum to 1, it stabilizes the estimators and therefore has a smaller variance than the original form. Theses two reasons make it not surprising for the Horvitz-Thompson ratio estimator to perform best among the four estimators.\\

\begin{figure}[H]
\centering
\begin{tabular}{c c}
\includegraphics[scale=0.4]{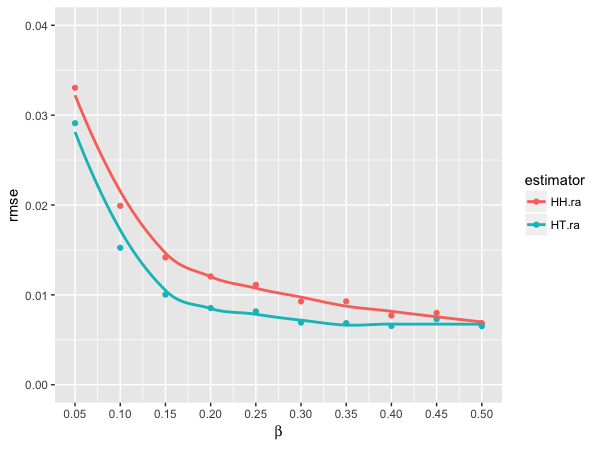}  
& \includegraphics[scale=0.4]{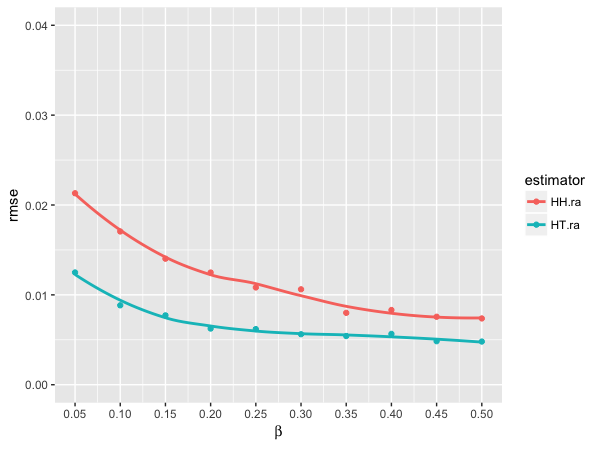}\\
(a) $n=5000$, $c.v.=2.4$, $\beta=0.2$  & (b) $n=5000$, $c.v.= 0.8$, $\beta=0.2$, $\gamma = 0.3$
\end{tabular}
\caption{Performance (measured by RMSE) of generalized Hansen-Hurwitz ratio estimator and Horvitz-Thompson ratio estimator versus number ($H$) of random walks in networks, measured by $MAD$, $RMSE$, and $KL$ (low values are better).}
\label{HH_HT_length}
\end{figure}

In Figure \ref{HH_HT_length}, we compare the performance of the Horvitz-Thompson ratio estimator and the generalized Hansen-Hurwitz ratio estimator by plotting their RMSE versus the sampling budget $\beta$. As one can observe, for the Horvitz-Thompson ratio estimator, we can use a smaller sampling budget to achieve the same estimation precision as the generalized Hansen-Hurwitz ratio estimator. For example, in network $(a)$, the estimation precision by the generalized Hansen-Hurwitz ratio estimator with $20\%$ sampling budget can be achieved by the Horvitz-Thompson ratio estimator with only about $12.5\%$ sampling budget. Therefore in practice, people would prefer using the Horvitz-Thompson ratio estimator if sampling is expensive and saving sampling budget is needed.

\subsection{Evaluation of Estimation}
In order to evaluate how well our estimates from section 4.4  perform in estimating the population SPLD, we first compare the generalized Hansen-Hurwitz ratio estimates, denoted by HH.ra, to the unweighted sample SPLDs observed from the induced subgraphs, denoted by UW. Note that by using HH.ra, we are correcting bias from UW, but the bias to be corrected for networks with large $c.v.$ and networks with small $c.v$ are different. For networks with large $c.v.$, we only correct the bias from unequal sampling probabilities, because we are still using the observed SPLs between sampled nodes from the induced subgraph. For networks with small $c.v.$, we correct bias from both unequal sampling probabilities and not observing the true SPLs between sampled nodes, as we use landmarks to estimate those SPLs.\\

\begin{figure}[H]
\centering
\begin{tabular}{cc}
\includegraphics[scale=0.4]{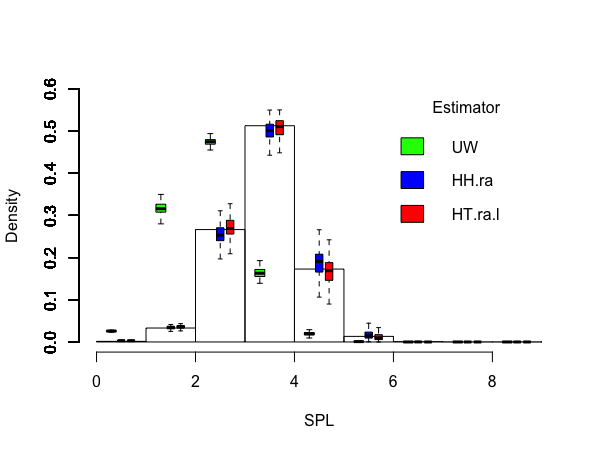} &
\includegraphics[scale=0.4]{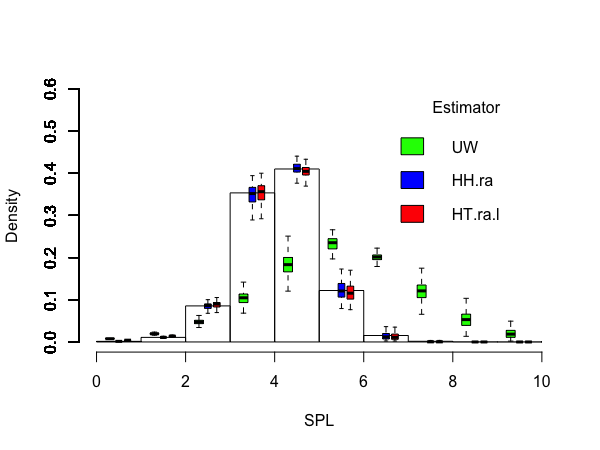} \\
(a) $n=5000$, $c.v.=2.4$, $\beta=0.2$ &
(b) $n=5000$, $c.v.=0.8$, $\beta=0.2$, $\gamma=0.3$
\end{tabular}
\caption{Box plots comparison of generalized Hansen-Hurwitz ratio estimates based on approximated SPL (HH.ra), unweighted sample SPLD observed from the induced subgraphs (UW), and generalized Hansen-Hurwitz ratio estimates based on actual SPL (HH.ra.l)}
\label{evaluation_boxplots}
\end{figure}

From the numerical comparison of UW and HH.ra in Table \ref{evaluation_numerical}, one can observe that for both networks, about $90\%$ of the estimation error in UW is reduced by using HH.ra. In Figure \ref{evaluation_boxplots}, one can observe that for networks with large $c.v.$ as shown in $(a)$, the box plots for UW are shifted to the left of the population SPLD. This is because dyads with shorter SPLs are more likely to the be sampled than dyads with longer SPLs, and thus the fractions of dyads with shorter SPLs are over estimated while the fractions of dyads with long SPLs are under estimated. Therefore for networks with large $c.v.$, bias from unequal sampling probabilities is dominating in the estimation error of UW. For networks with small $c.v.$ as shown in $(b)$, the box plots for UW are shifted to the right of the population SPLD. This is because the many observed SPLs are longer than the true SPLs. Therefore in networks with small $c.v.$, bias from not observing the true SPLs between sampled nodes is dominating in the estimation error of UW, and thus correcting it is necessary. For both networks, after applying HH.ra, the box plots stay at the right positions on the histogram with short whisker, which means the estimates are unbiased and have small variance.\\

On the other hand, in order to see how much we can improve if we can actually observe the true SPLs between sampled nodes, we compare  our HH.ra based on approximated SPLs, to the generalized Hansen-Hurwitz estimates based on the true SPLs between sampled nodes, denoted by HH.ra.l. As we can observe, there will still be some improvement if we use the latter, but the improvement will not be huge. More specifically, in Table \ref{evaluation_numerical}, the improvement from HH.ra to HH.ra.l is only about $10\%$. In Figure \ref{evaluation_boxplots}, we can also see that the box plots for HH.ra and those for HH.ra.l are really close. Therefore in practice, we will prefer to base our estimates on the approximated SPLs for saving computation time and not loosing much estimation accuracy. 

\begin{table}[H]
\begin{center}
\begin{tabular}{c | c c c |c c c}
 & & $n=5000$, $c.v.=2.4$ & & & $n=5000$, $c.v.=0.8$&\\
 & & $\beta=0.2$ & & & $\beta=0.2$, $\gamma=0.3$&\\
 &MAD & RMSE & KL & MAD & RMSE & KL\\
\hline
Unweighted sample SPLD & .114 & .115 & 0.161 & .101 & .103 & 0.213\\
HH.ra by approximated SPL & .010 & .012 & .0025 & .011 & .014 & .0036\\
HH.ra by real SPL & .009 & .011 & .0023 & .010 & .013 & .0034
\end{tabular}
\end{center}
\caption{Numerical comparison of generalized Hansen-Hurwitz ratio estimates based on approximated SPL (HH.ra), unweighted sample SPLD observed from the induced subgraphs (UW), and generalized Hansen-Hurwitz ratio estimates based on actual SPL (HH.ra.l).}
\label{evaluation_numerical}
\end{table}

\section{Applications}

\begin{figure}[H]
\centering
\begin{tabular}{c c}
\includegraphics[scale=0.3]{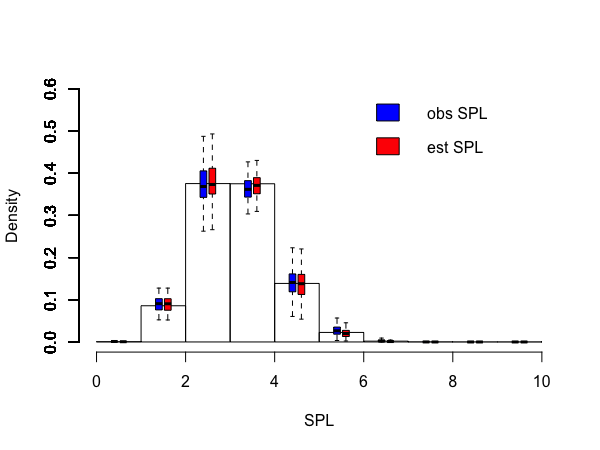}  
& \includegraphics[scale=0.3]{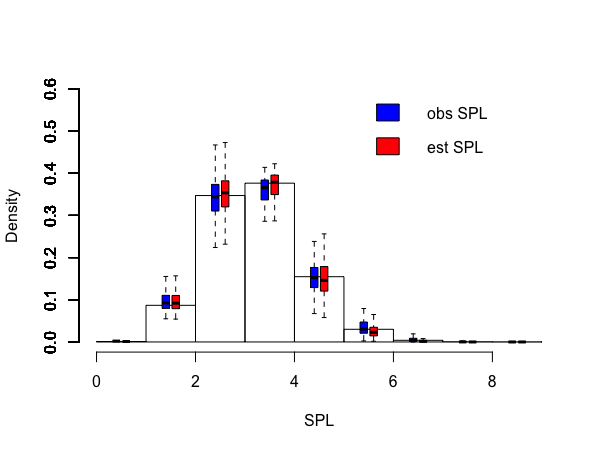}\\
(a) Oregon ($c.v. = 7.6$) & (b) AS-733 ($c.v. = 5.8$)\\
\includegraphics[scale=0.3]{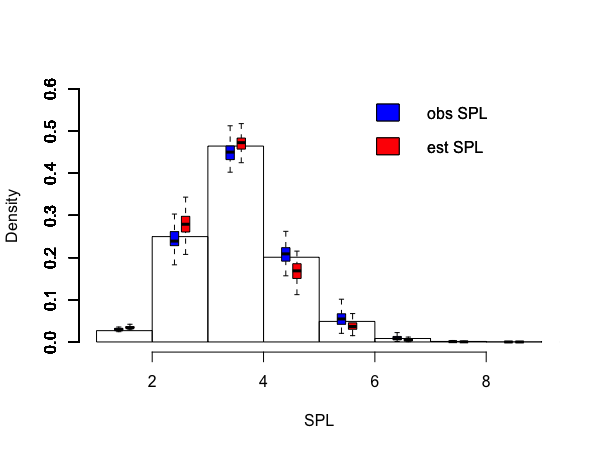}  
& \includegraphics[scale=0.3]{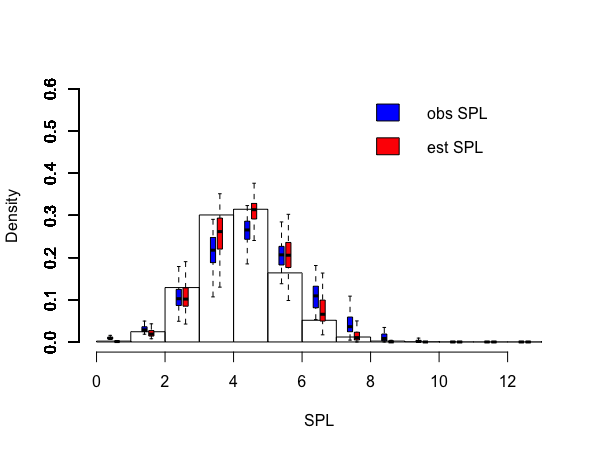}\\
(c) Email-Enron ($c.v. = 3.5$) & (d) CA-HepPh ($c.v. = 2.29$)\\
\includegraphics[scale=0.3]{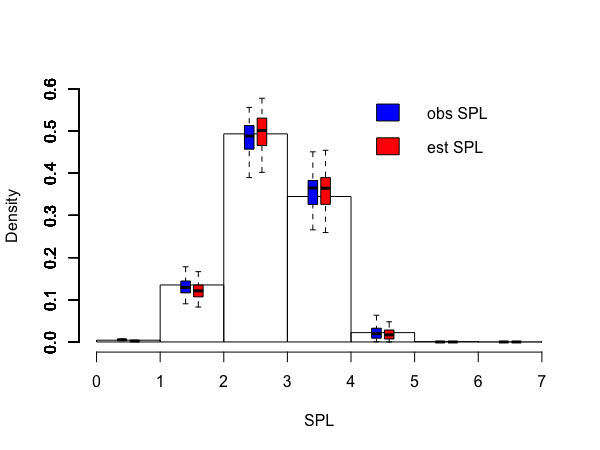}  
& \includegraphics[scale=0.3]{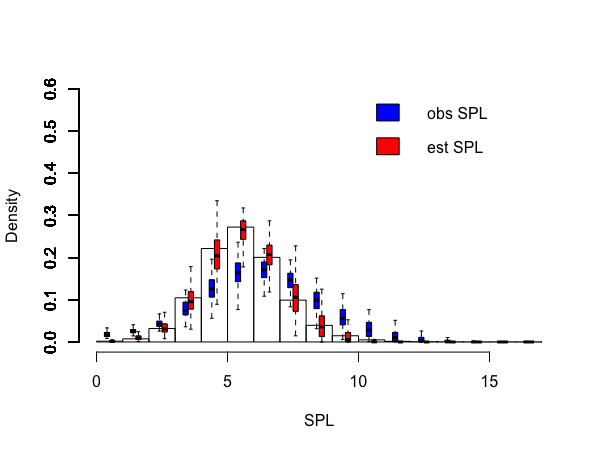}\\
(e) Wiki-Vote ($c.v. = 2.06$) & (f) CA-HepTh ($c.v. = 1.12$)\\
\includegraphics[scale=0.3]{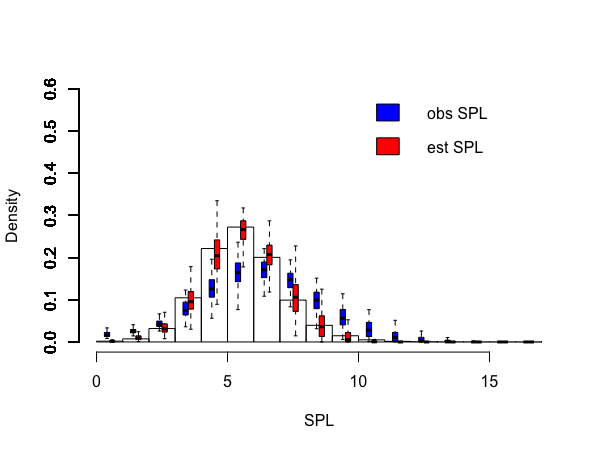}  
& \includegraphics[scale=0.3]{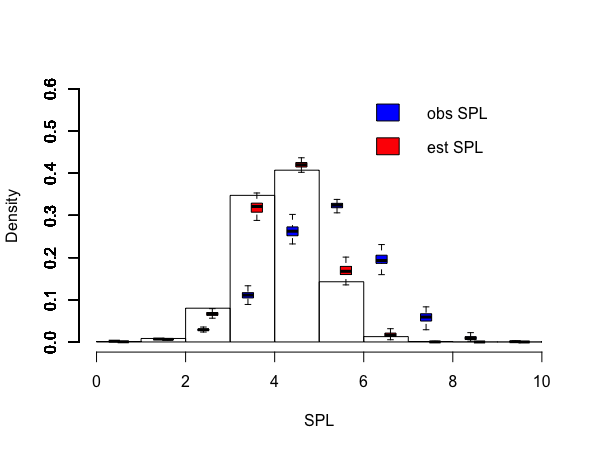}\\
(g) CA-GrQc ($c.v. = 1.34$) & (h) P2P ($c.v. = 0.9$)\\
\end{tabular}
\caption{Box plots of estimated SPLDs of real networks: HH.ra by observed SPL ($\beta=0.2$) v.s. HH.ra by estimated SPL by landmarks ($\beta=0.2$, $\gamma=0.3$).}
\label{real_boxplots}
\end{figure}

In this section, we test our SPLD estimation methods on data from eight real world networks. These data are available on the SNAP (Stanford Network Analysis Project) website. To simplify the analysis, we only consider nodes in the largest connected component. Table \ref{summary} summarizes the basic information for each network used in out test. These networks vary in size, number of edges, average degree, and most importantly, coefficient of variation. We compare the HH.ra estimates based on observed SPLs from induced subgraph (obs SPL) versus those based on estimated SPLs by landmarks (est SPL). For estimates based on observed SPLs from induced subgraph, we use a single random walk with $20\%$ sampling budget. For estimates  based on estimated SPLs by landmarks, we use  a single random walk with $20\%$ sampling budget and $30\%$ of the sampled nodes as landmarks. The results are shown in Table \ref{real_numerical} and Figure \ref{real_boxplots}.\\

As shown in the first $(a)$, $(b)$, and $(c)$ in Figure \ref{real_boxplots}, the estimates based on observed SPLs of the first three real networks, Oregon, AS-733, and Email-Enron are very good. This is not surprising, as the $c.v.$'s for those networks are all much larger than 2, which indicates the existence of hubs. In addition, the performance on Email-Enron network is the best among these three, as measured by the small values in MAD, RMSE, and KL in Table Table \ref{real_numerical}. This is also to be expected, since Email-Enron network has the largest size among the three. According to our discussion in section 5.3, our estimation method tends to perform better for larger networks.\\

When the $c.v.$ gets closer to 2, the performance of estimation based on observed SPLs varies from case to case. For example, Figure \ref{real_boxplots} $(c)$ and $(d)$ show that the SPLD of CA-HepPh network is a little over-estimated, while the SPLD of Wiki-Vote network is very well estimated. As one can observe, the average distance in CA-HepPh network is longer than the average distance in Wiki-Vote network, therefore random walks in CA-HepPh network are having a harder time in finding the true shortest paths. The performance of estimation is getting worse as the $c.v.$ decreases to some value below 1.5, and even below 1. For networks CA-HepTh, CA-GrQc, and P2P, the SPLDs are highly over estimated. The worst case happens to the P2P network, which only has $c.v.=0.9$. Since there's no powerful hub in networks $(f)$, $(g)$, and $(h)$ , its really hard for random walks to find the shortest paths.\\

Alternatively, we can base the estimates on the estimated SPLs by landmarks. As one can notice, for networks whose estimates based on the observed SPLs are good, such as $(a)$, $(b)$, $(c)$, and $(e)$, there won't be much improvement if we base the estimates on the estimated SPLs. However, for networks with small value in $c.v.$, whose estimates based on the observed SPLs are far from the true SPLDs, such as $(f)$, $(g)$, and $(h)$, using estimated SPLs will correct the bias from not observing true SPLs in the induced subgraph and therefore result in much better estimation performance.

\begin{table}[H]
\begin{center}
\begin{tabular}{c | c c c c c}
Network & nodes & edges & $<k>$ & $cv$ & $E.f$\\
\hline
Oregon & 10.7K & 22K & 4.1 & 7.6 & 0.162\\
AS-733 & 6.4K & 13.2K & 4.3 & 5.8 & 0.140\\
Email-Enron & 33.7K & 361.7K & 21.5 & 3.5 & 0.298\\
CA-HepPh & 11.2K & 235.2K & 42 & 2.29 & 0.361\\
Wiki-Vote & 7.1K & 103.7K & 29.3 & 2.06 & 0.254\\
CA-HepTh & 8.6K & 49.6K & 11.5 & 1.12 & 0.107\\
CA-GrQc & 4.2K & 26.8K &  12.9 & 1.34 & 0.129\\
P2P & 10.9K & 40K & 7.4& 0.9 & 0.093\\
\end{tabular}
\end{center}
\caption{Summary of Networks}
\label{summary}
\end{table}

\begin{table}[H]
\begin{center}
\begin{tabular}{c c | c c c }
& HH.ra by & MAD & RMSE & KL\\
\hline
Oregon
& obs SPL & .012 & .014 & .0032 \\
& est SPL & .011 & .014 & .0029\\
\hline
AS-733
& obs SPL & .016 & .021 & .0055 \\
& est SPL & .016 & .020 & .0051\\
\hline
Email-Enron
& obs SPL & .0069 & .009 & .0023 \\
& est SPL  & .0085 & .010 & .0032\\
\hline
CA-HepPh
& obs SPL & .026 & .032 & .026 \\
& est SPL & .016 & .022 & .011\\ 
\hline
Wiki-Vote
& obs SPL & .014 & .018 & .0028 \\
& est SPL & .015 & .018 & .0029\\
\hline
CA-HepTh
& obs SPL & .028 & .034 & .054 \\
& est SPL & .010 &  .015 & .012\\
\hline
CA-GrQc
& obs SPL & .031 & .038 & .062 \\
& est SPL & .015 & .024 & .0225\\
\hline
P2P
& obs SPL & .086 & .087 & .13 \\
& est SPL & .009 & .010 & .0012
\end{tabular}
\end{center}
\caption{Numerical evaluation measures of estimated SPLDs of real networks: HH.ra by observed SPL ($\beta=0.2$) v.s. HH.ra by estimated SPL by landmarks ($\beta=0.2$, $\gamma=0.3$).}
\label{real_numerical}
\end{table}

\printbibliography

\end{document}